\documentclass[12pt,preprint]{aastex}

\newcommand{\fb}{f_{b}}
\newcommand{\kmax}{k_{R_C\rightarrow B}^{\rm{max}}}
\newcommand{\krtob}{k_{R_C\rightarrow B}}
\newcommand{\vmax}{v^{\rm{max}}}

\newcommand{\rpmin}{r_p^{\rm{min}}}
\newcommand{\rpmax}{r_p^{\rm{max}}}

\newcommand{\frem}{f_{{\rm rem}}}
\newcommand{\fmg}{f_{{\rm mg}}}
\newcommand{\tmg}{T_{{\rm mg}}}

\newcommand{\arcsecpt}{\hbox to 1pt{}\rlap{\arcsec}.\hbox to 2pt{}}
\newcommand{\arcminpt}{\hbox to 1pt{}\rlap{\arcmin}.\hbox to 2pt{}}

\newcommand{\hkpc}{$h^{-1}$~kpc}
\newcommand{\hmpc}{$h^{-1}$~Mpc}

\newcommand{\mlim}{M_{\rm{lim}}}
\newcommand{\rclim}{R_{C_{\rm{lim}}}}
\newcommand{\lsun}{L_{\odot}}

\slugcomment{Accepted for Publication in the Astrophysical Journal}

\shorttitle{CNOC2 Close Pairs and Merger Rate Evolution}
\shortauthors{Patton et al.}

\begin{document}

%% LaTeX will automatically break titles if they run longer than
%% one line. However, you may use \\ to force a line break if
%% you desire.

\title{Dynamically Close Galaxy Pairs and Merger Rate Evolution\\ 
in the CNOC2 Redshift Survey}

%% Use \author, \affil, and the \and command to format
%% author and affiliation information.
%% Note that \email has replaced the old \authoremail command
%% from AASTeX v4.0. You can use \email to mark an email address
%% anywhere in the paper, not just in the front matter.
%% As in the title, you can use \\ to force line breaks.

\author{D. R. Patton\altaffilmark{1,2,3,12}, 
C. J. Pritchet\altaffilmark{2,12}, 
R. G. Carlberg\altaffilmark{3,12},
R. O. Marzke\altaffilmark{4,11,12},
H. K. C. Yee\altaffilmark{3,12}, 
P. B. Hall\altaffilmark{3,5,12},
H. Lin\altaffilmark{3,6,11,12},
S. L. Morris\altaffilmark{7,8,10},
M. Sawicki\altaffilmark{3,9,12},
C. W. Shepherd\altaffilmark{3},
\& 
G. D. Wirth\altaffilmark{10,12}
}
\altaffiltext{1}{Department of Physics, Trent University, 
1600 West Bank Drive, Peterborough, ON, K9J 7B8, Canada
\email{dpatton@trentu.ca}}
\altaffiltext{2}{Department of Physics and Astronomy, 
University of Victoria,
PO Box 3055, Victoria, BC, V8W 3P6, Canada 
%\email{pritchet@uvastro.phys.uvic.ca}
}
\altaffiltext{3}{Department of Astronomy and Astrophysics, 
University of Toronto, 
60 St. George Street, Toronto, ON, 
M5S 3H8, Canada 
%\email{carlberg, hyee, hall, shepherd@astro.utoronto.ca}
}
\altaffiltext{4}{Carnegie Observatories, 813 Santa Barbara St.,
Pasadena, CA 91101 
%\email{marzke@ociw.edu}
}
\altaffiltext{5}{Princeton University Observatory, Princeton, NJ, 
08544-1001 and Pontificia Universidad Cat\'{o}lica de Chile, 
Departamento de Astronomia y Astrofisica, Facultad de Fisica, 
Casilla 306, Santiago 22, Chile}
\altaffiltext{6}{Steward Observatory, University of Arizona, 933
N. Cherry Avenue, Tucson, AZ 85721 
%\email{hlin@as.arizona.edu}
}
\altaffiltext{7}{Department of Physics, University of Durham, 
South Road, Durham DH1 3LE, UK 
%\email{simon.morris@durham.ac.uk}
}
\altaffiltext{8}{Dominion Astrophysical Observatory, Herzberg Institute
of Astrophysics, Victoria, BC V8X 4M6, Canada}
\altaffiltext{9}{California Institute of Technology, 320-47, 
Pasadena, CA 91125 
%\email{sawicki@mop.caltech.edu}
}
\altaffiltext{10}{W. M. Keck Observatory, Kamuela, HI 96743
%\email{wirth@keck.hawaii.edu}
}
\altaffiltext{11}{Hubble Fellow}
\altaffiltext{12}{Visiting Astronomer, Canada-France-Hawaii 
Telescope, which is operated
by the National Research Council of Canada, le Centre National 
de Recherche Scientifique, and the University of Hawaii.}

%% Notice that each of these authors has alternate affiliations, which
%% are identified by the \altaffilmark after each name.  Specify alternate
%% affiliation information with \altaffiltext, with one command per each
%% affiliation.

%% Mark off your abstract in the ``abstract'' environment. In the manuscript
%% style, abstract will output a Received/Accepted line after the
%% title and affiliation information. No date will appear since the author
%% does not have this information. The dates will be filled in by the
%% editorial office after submission.

\begin{abstract}
We investigate redshift 
evolution in the galaxy merger and accretion rates, 
using a well-defined sample of 4184 galaxies with 
$0.12 \leq z \leq 0.55$ and $R_C \leq 21.5$.   
We identify 88 galaxies in close ($5\leq r_p \leq 20~$\hkpc) 
dynamical ($\Delta v \leq 500$ km/s) pairs. 
These galaxies are used to compute global pair statistics,  
after accounting for selection effects resulting from 
the flux limit, $k$-corrections, 
luminosity evolution, and spectroscopic incompleteness.
We find that the number of companions per galaxy 
(for $-21 \leq M_B^{k,e} \leq -18$)
is $N_c = 0.0321\pm 0.0077$ at $z$=0.3.  
The luminosity in companions, per galaxy, is 
$L_c = 0.0294\pm 0.0084 \times 10^{10}~h^2 \lsun$.  
We assume that $N_c$ is proportional to the galaxy merger rate, 
while $L_c$ is directly related to the mass accretion rate.
After increasing the maximum pair separation to 50 \hkpc,
and comparing with the low redshift SSRS2 pairs sample, 
we infer evolution in the galaxy merger and accretion rates
of $(1+z)^{2.3 \pm 0.7}$ and $(1+z)^{2.3 \pm 0.9}$
respectively.  These are the first such estimates to be 
made using only confirmed dynamical pairs.
When combined with several additional assumptions, 
this implies that approximately
15\% of present epoch galaxies with $-21 \leq M_B \leq -18$ 
have undergone a major merger since $z$=1.  
\end{abstract}

%% Keywords should appear after the \end{abstract} command. The uncommented
%% example has been keyed in ApJ style. See the instructions to authors
%% for the journal to which you are submitting your paper to determine
%% what keyword punctuation is appropriate.

\keywords{galaxies : evolution --- galaxies : interactions --- surveys}  

%% From the front matter, we move on to the body of the paper.
%% In the first two sections, notice the use of the natbib \citep
%% and \citet commands to identify citations.  The citations are
%% tied to the reference list via symbolic KEYs. The KEY corresponds
%% to the KEY in the \bibitem in the reference list below. We have
%% chosen the first three characters of the first author's name plus
%% the last two numeral of the year of publication as our KEY for
%% each reference.

\section{INTRODUCTION} \label{cnoc2mr:intro}

The nearby universe contains a number of striking examples of galaxies 
which are in the process of merging with one another.  While this 
phenomenon is relatively rare, it can lead to significant changes 
in the structure and stellar makeup of the galaxies involved.  
Considerable effort has been invested in modelling the dynamics of 
various merging systems (e.g., 
Barnes 1988\nocite{B88}) and in measuring 
the effects of close encounters at various wavelengths (e.g., 
Sanders \& Mirabel 1996\nocite{SM96}).  
However, it remains unclear how 
important this process is for galaxies in general.  In order to 
better understand the role that mergers play in the evolution of 
galaxy populations, one must determine the timescale 
of these events and the rate at which they occur at different epochs.  
Large redshift surveys of galaxies at cosmologically significant 
lookback times have now made it feasible to measure changes in 
the merger rate with redshift.

From an observational standpoint, mergers can be studied
at different stages in the merger process, ranging from well separated 
galaxy pairs (early stage mergers) to strongly interacting systems 
(e.g., the Antennae) and late stage mergers (e.g., Arp 220).
Exotic phenomena such as ultra-luminous infrared galaxies (ULIRG's), 
active galactic nuclei (AGN), 
radio galaxies, shell galaxies, and ring galaxies have been 
associated with some of these different stages 
(see Struck 1999\nocite{struck} for a review).  
For this study, we will focus on close pairs of galaxies, which 
are good candidates for early stage mergers if chosen carefully.
With appropriate selection criteria, these systems can be identified 
in a relatively straightforward and objective manner.

There have been a number of studies in which close pairs of 
galaxies have been used to measure evolution in the galaxy merger 
rate \citep{ZK,BKWF,CPI,YE,WOODS,P97,CFRS00}.  
These studies have yielded a 
wide variety of results.  Some of the disparity has been 
attributed to differences in pair definitions and techniques, 
and in most cases, the error bars have been quite large.
Nevertheless, significant discrepancies remain.  Patton et al. (2000; 
hereafter P2000)\nocite{ssrs2mr} 
demonstrated that some differences can be attributed to the 
fact that pair statistics (e.g., the fraction of galaxies in pairs) depend 
sensitively on the survey {\it depth}.  That is, even for a simple 
volume-limited redshift survey, pair statistics will naturally 
increase with depth.  For the same reason, pair statistics 
computed for a flux-limited redshift survey will have an 
unwanted redshift-dependent selection bias.   Moreover, {\it any} 
selection effect which changes the observed mean density of galaxies 
in the sample (e.g., spectroscopic incompleteness) will have a similar 
effect on pair statistics.  Without correction, these biases can 
hinder comparison of different surveys and lead to significant 
- and potentially severe - biases in merger rate estimates.

To account for these biases,
P2000\nocite{ssrs2mr} introduced two new 
pair statistics, and outlined methods for accounting for 
differences in sample depth and completeness.  These methods 
were tested extensively, and 
were shown to yield robust comparisons of 
pair statistics at different redshifts.   
This approach was then applied to the large, 
well-defined SSRS2 survey \citep{SSRS2}, yielding the first secure 
estimates of pair statistics at low redshift ($z\sim 0$).  
These measurements indicate that roughly 2\% of galaxies with 
$-21 \leq M_B \leq -18$ are found in close ($5\leq r_p \leq 20~$\hkpc) 
dynamical ($\Delta v \leq 500$ km/s) pairs. 

In this study, we apply these techniques to 
the CNOC2 Field Galaxy Redshift Survey 
(hereafter CNOC2; Yee et al. 2000\nocite{CNOC2data}).  
This large, well-defined sample of galaxies at moderate 
redshift ($0.12 < z < 0.55$) will be used to 
establish how pair statistics evolve with 
redshift.  This will allow us to infer changes in 
the galaxy merger and accretion rates.  
\subsection{Survey Overview}
The CNOC2 survey covers 4 well-separated 
patches on the sky, each 
subtending $\sim 0.4~{\rm deg}^2$.  These patches have been 
assigned names based on their equatorial coordinates (B1950.0), 
as follows : 0223+00, 0920+37, 1447+09, and 2148-05.  
Each patch consists of a contiguous L-shaped region with a 
central block.
These patches were chosen to avoid bright stars, known 
low-redshift clusters, and other bright objects at low redshift.
Data were acquired during 7 observing runs at the 
Canada-France-Hawaii Telescope (CFHT), between February 1995 
and May 1998.  All imaging and spectroscopic data were obtained 
with CFHT's 
Multi-Object Spectrograph (MOS).  A total of 74 MOS fields 
were observed, each of size 
$\sim 9\arcmin \times 8\arcmin$.

In Section~\ref{cnoc2mr:data}, 
we describe the relevant aspects of the CNOC2 survey.
An overview of the basic methods for computing pair 
statistics is given in Section~\ref{cnoc2mr:methods}.  
We then discuss how galaxies are selected 
for the primary and secondary samples required for the 
pairs analysis.  A detailed treatment of selection effects 
is given in \S~\ref{cnoc2mr:select}, accounting for 
flux limits, luminosity evolution, spectroscopic incompleteness, 
and boundary effects.  CNOC2 pair statistics are presented 
in Section~\ref{cnoc2mr:nclc}, and used to measure evolution 
in the galaxy merger and accretion rates (\S~\ref{cnoc2mr:mrate}).  
Results are discussed in the final section.  
Unless indicated otherwise, we adopt 
$H_0$=100 km s$^{-1}$Mpc$^{-1}$ ($h$=1) and $q_0$=0.1.

\section{The CNOC2 Survey} \label{cnoc2mr:data}
The CNOC2 survey consists of redshifts for~$\sim$~5000 field galaxies 
spanning the redshift range $0.1 < z < 0.6$.  
A detailed description of the CNOC2 observing and data 
reduction methods, along with the first of four data 
catalogs\footnote{This paper uses the most current versions of the 
CNOC2 catalogs: v1\_9V for the 0223+00 patch, 
and v1\_8V for the other three patches.}, 
is given by Yee et al.\nocite{CNOC2data} (2000).  
Here, we give a brief overview of the 
survey, focussing on aspects relevant to this study. 

\subsection{Survey Overview}
The CNOC2 survey covers 4 well-separated 
patches on the sky, each 
subtending $\sim 0.4~{\rm deg}^2$.  These patches have been 
assigned names based on their equatorial coordinates (B1950.0), 
as follows : 0223+00, 0920+37, 1447+09, and 2148-05.  
Each patch consists of a contiguous L-shaped region with a 
central block.
These patches were chosen to avoid bright stars, known 
low-redshift clusters, and other bright objects at low redshift.
Data were acquired during 7 observing runs at the 
Canada-France-Hawaii Telescope (CFHT), between February 1995 
and May 1998.  All imaging and spectroscopic data were obtained 
with CFHT's 
Multi-Object Spectrograph (MOS).  A total of 74 MOS fields 
were observed, each of size 
$\sim 9\arcmin \times 8\arcmin$.   

\subsection{Photometry}\label{cnoc2mr:phot}
Images of all patches were obtained in Kron-Cousins $R_C$ 
and $I_C$, and 
Johnson $U$, $B$, and $V$, using MOS in imaging mode.  
Exposure times ranged from 6 to 15 minutes.  
Object detection, star-galaxy classification, and photometry 
were carried out using an improved 
version of the Picture Processing Package \citep{PPP,YEC}. 
We correct our photometry for extinction 
from the Milky Way (see Lin et al. 1999\nocite{CNOC2LF}).  
In this study, we will use 
observations in $R_C$ and $B$, which have 
average 5$\sigma$ detection limits of 24.0 and 24.6 
respectively. 
The primary spectroscopic sample is chosen in the $R_C$ band, 
and we adopt $R_C$=21.5 as the nominal spectroscopic 
completeness limit.  

\subsection{Spectroscopy}\label{cnoc2mr:spectra}
Spectra were obtained using the B300 grism, providing 
a resolution of $\sim 15$\AA.  A band-limiting filter 
was used to enable stacking of spectra, increasing 
the size of the redshift sample.  The wavelength coverage 
of this filter is $4400$\AA~ to $6300$\AA.  This allows 
for the identification of important spectral features 
in galaxies of all spectral types, over the redshift range 
$0.12 \leq z \leq 0.55$.  
Redshift measurements were performed using cross-correlation 
techniques, yielding rms velocity errors of approximately 
100 km/s.  

For reasons of observational efficiency, we did not attempt 
to obtain spectra for the complete sample of galaxies 
with $R_C\leq 21.5$.
Instead, two multi-slit masks were used for each field, yielding 
a total of 80-90 redshifts in most cases.  
The cumulative redshift 
sampling rate, defined as the fraction of $R_C\leq 21.5$ 
galaxies with measured redshifts, is about 50\%.  The differential 
redshift sampling rate, which gives the sampling rate at 
a given apparent magnitude, is highest at bright magnitudes, 
and decreases to $\sim 20\%$ at $R_C$=21.5.  We have carefully 
accounted for selection effects that result from the 
spectroscopic incompleteness of our sample.  
Discussion of these selection weights is deferred to 
Sections~\ref{wmc} and \ref{wtheta}.  

\subsection{The $B$-band Luminosity Function} \label{cnoc2mr:lf}

In order to extract meaningful pair statistics from a flux-limited sample, 
it is necessary to correct the statistics to a specified range in 
absolute magnitude (P2000\nocite{ssrs2mr}).  
We will use the observed galaxy luminosity function (LF)
to make this correction.
\citet{CNOC2LF} have carried out a detailed 
study of the CNOC2 LF, introducing a convenient parameterization 
of luminosity and number density evolution.  
Here, we summarize their results briefly.  

The differential galaxy LF, denoted $\phi(M)$, gives the 
co-moving number density of galaxies of absolute magnitude $M$.  
The usual \citet{SCH76} 
parameterization of this function was adopted,    
with characteristic absolute magnitude $M^*$, faint-end slope 
$\alpha$, and normalization $\phi ^*$.  
Numerous studies have revealed significant evolution in the 
galaxy LF, even at the fairly modest redshifts of concern 
here (see Ellis 1997\nocite{ELLIS97} for a review).  Thus, the LF must be 
generalized to the form $\phi(M,z)$.  
The redshift dependence was parameterized as follows : 
\begin{eqnarray}
M^*(z) & = & M^*(0.3) - Q (z-0.3) \nonumber \\ 
\alpha(z) & = & \alpha(0) \label{eqmodel} \\
\phi^*(z) & = & \phi^*(0) 10^{0.4 P z} \nonumber \ .
\end{eqnarray} 
Here, $Q$ provides a linear fit to $M^*$ or luminosity evolution, 
while $P$ models density evolution.
Five color photometry ($UBVR_CI_C$) was used 
to classify CNOC2 galaxies according to their spectral energy 
distributions (SED's).
Treating early, intermediate, and late-type galaxies separately, 
the galaxy LF was computed in the rest-frame $B_{AB}$, $R_C$, 
and $U$ bandpasses.  We will use the $B_{AB}$ LF parameters given 
in Table 1 of \citet{CNOC2LF}.
As we are concerned only with the global field population, 
we sum up LF's for the three SED types.  In order to 
compare with $B$-band pair statistics at low redshift, we 
transform from $B_{AB}$ to $B$, using the relation $B = B_{AB} + 0.14$ 
\citep{FSI95}.

\section{BASIC METHOD} \label{cnoc2mr:methods}
This study relies heavily on the techniques introduced by 
P2000\nocite{ssrs2mr} 
for measuring pair statistics.  By extending their 
methods to higher redshift, and making a direct comparison 
with their measurements at $z \sim 0$, we will investigate 
redshift evolution in pair statistics.  
In this section, we provide a brief summary of the methods and 
results of P2000\nocite{ssrs2mr} that will be used in this paper.

\subsection{Problems With Traditional Close Pair Statistics}

Close pairs of galaxies provide the best available means 
of estimating the galaxy merger rate, and its evolution with 
time.  Close pair statistics provide 
an integrated measure of galaxy clustering on small 
scales, and are often assumed to be 
largely independent of selection effects such as 
sampling depth and completeness.  
However, measurements of pair statistics, like the merger rate itself, 
necessarily depend both on clustering {\it and} the mean number density 
of galaxies in the sample.  
In order to account for the latter,
one must identify a specific range in 
absolute magnitude (or mass) when computing pair statistics, if the 
results are to be physically meaningful. 
Furthermore, when comparing the pair statistics of 
different samples, one must ensure that identical ranges in absolute 
magnitude (or mass) are used for all samples.

When applying pair statistics to flux-limited surveys, 
it is necessary to select a specific range in absolute magnitude 
which is representative of the sample.  One must then 
correct the pair statistics for changes in sampling 
depth with redshift.  
A number of pair statistics are not well suited to 
making these corrections.  This includes the observed fraction 
of galaxies in pairs, and nearest neighbour statistics.  
Statistics which use the number or luminosity of companions
are the most straightforward to apply to a flux-limited 
redshift survey.  

\subsection{Two New Pair Statistics}\label{cnoc2mr:newstats}

Two robust pair statistics were introduced.
The number of close companions per galaxy, hereafter 
called $N_c$, is directly related to the galaxy merger rate.  
The luminosity in companions per galaxy, designated $L_c$, 
is related instead to the mass accretion rate.  
These statistics are measured using primary and secondary samples 
of galaxies, where one searches for companions (from the 
secondary sample) close to host galaxies (primary sample).  
A weighting scheme was introduced, allowing meaningful statistics 
to be computed for a flux-limited sample.  
These weights recover correctly the equivalent volume-limited pair 
statistics (verified with Monte Carlo simulations) and 
minimize the uncertainty in the measured pair statistics.  

\subsection{A Useful Definition of a Close Pair}

Pairs in redshift space can be uniquely specified by 
their projected physical separation ($r_p$) and rest-frame 
line-of-sight velocity difference ($\Delta v$).  
A useful definition of a close companion that is likely 
to merge soon ($\tmg \lesssim 0.5$ Gyr) 
is 5~\hkpc$ < r_p \leq 20$ \hkpc~ and $\Delta v \leq 500$ km/s.  
These choices make sense from a theoretical standpoint; 
moreover, at least half of the SSRS2 pairs satisfying these criteria 
exhibit clear morphological signs of on-going interactions.  

\subsection{SSRS2 Close Pair Statistics}\label{ssrs2:nclc}

Pair statistics were estimated for the SSRS2 redshift survey, using 
the pair definition given above.  For galaxies with absolute magnitudes 
$-21 \leq M_B \leq -18$, P2000\nocite{ssrs2mr} found
$N_c=0.0226 \pm 0.0052$ and $L_c = 0.0216 \pm 0.0055 
\times 10^{10}~h^2 \lsun$ at $z$=0.015.

\section{SAMPLE SELECTION}\label{cnoc2mr:sample}
We now extend these techniques to the CNOC2 survey.  
In this section, 
we outline our approach for identifying a well-defined 
sample of galaxies to be used in the pairs analysis.  
We begin by defining our choice of a flux limit. 
We then impose further restrictions on the sample in order to 
minimize potential biases due to $k$-corrections, 
luminosity evolution, and luminosity-dependent clustering.

\subsection{Flux Limit}\label{cnoc2mr:flux}
The CNOC2 survey is primarily flux-limited in nature.
Galaxies were originally selected for follow-up spectroscopy 
based on their $R_C$ apparent magnitudes.  As described 
in Section~\ref{cnoc2mr:data}, the nominal spectroscopic 
completeness limit for CNOC2 is $R_C$=21.5.  We adopt this 
as our initial flux limit.  
We further restrict our 
sample to the secure redshift range $0.12 \leq z \leq 0.55$ 
(see \S~\ref{cnoc2mr:spectra}).

\subsection{Estimating $B$ Absolute Magnitudes}\label{cnoc2mr:babs}

While galaxies were selected based on their $R_C$ flux, 
we are primarily interested in their absolute 
magnitudes in rest-frame $B$.  The primary reason for this choice 
is to enable us to perform a direct comparison with the 
low-redshift $B$-band pair statistics from the SSRS2 
survey.  The CNOC2 $B$ photometry will be used to measure 
galaxy absolute magnitudes, for use in the 
pair statistics.  It is important to note that observed $R_C$ corresponds 
to rest-frame $B$ at $z\sim$0.4.  For a galaxy at redshift $z$, 
absolute magnitude in rest-frame $B$ is given by
\begin{equation}\label{eqn:absmb} 
M_B^{k,e} (z) = B - 5\log d_L(z) - 25 - k(z) - E(z),
\end{equation}
where $k(z)$ is the $k$-correction and $E(z)$ is the correction for 
luminosity evolution.  
$B-$band $k$-corrections were estimated by
first fitting our 5-color photometry ($UBVR_CI_C$) to the 
SED models of \citet{CWW}.  
After interpolating to obtain an SED type, the resulting SED 
was then used to derive the $k$-correction.  

Modelling of luminosity evolution is based on measurements of 
LF evolution within the CNOC2 sample \citep{CNOC2LF}.  
A reasonable parameterization of luminosity evolution within 
the sample as a whole is a brightening of $Qz$ magnitudes for a 
galaxy at redshift $z$, with $Q$=1.  This choice of $E(z)$=$-z$ 
will be adopted through the remaining analysis,   
and its effects on the main results 
of this study will be discussed in Section~\ref{cnoc2mr:Qdepend}.  

\begin{figure}
\plotone{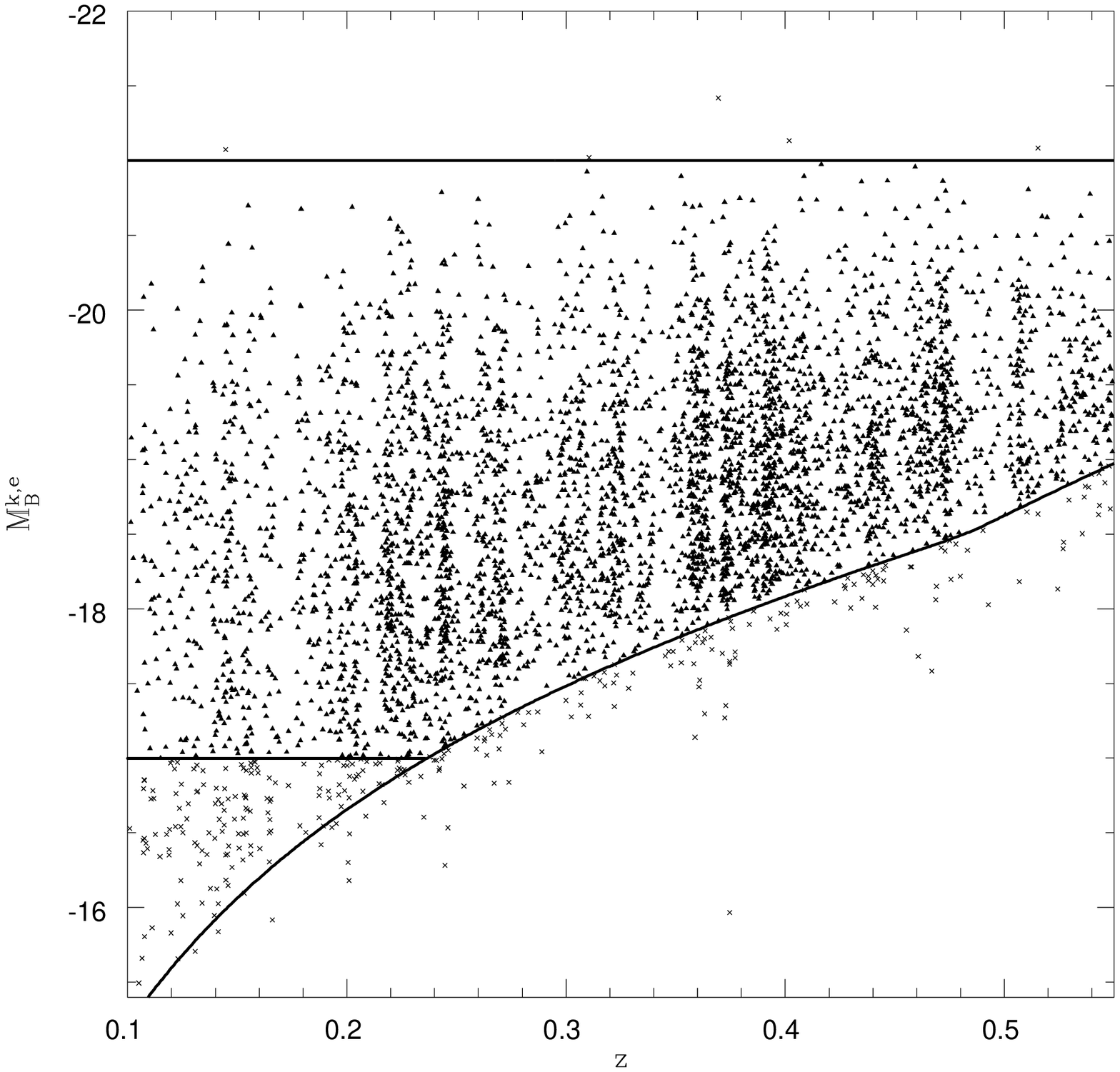}
\caption[$M_B-$Redshift Diagram and Sample Selection]
{$B$ absolute magnitude is plotted versus redshift for 
all CNOC2 galaxies with 
$R_C \leq 21.5$ and $0.12 \leq z \leq 0.55$. 
The curved line gives the limit imposed by the limiting 
apparent magnitude, assuming the maximal 
$k$-correction $\kmax (z)$ at redshift $z$.  
Note that objects lying below this line are brighter than 
$R_C$=$21.5$ but are blue; Section~\ref{cnoc2mr:limits} explains why these 
objects are excluded from the sample.
The upper horizontal line indicates the bright limit imposed 
on the sample($M_B^{k,e}$ = $-21$), while the lower horizontal line 
denotes the faint limit ($M_B^{k,e}$ = $-17$).  
Galaxies satisfying all of these criteria (and hence used 
in the calculation of $N_c$ and $L_c$) are marked 
with triangles; the remainder are indicated with crosses.
\label{cnoc2mr:figmbz}}
\end{figure}

In Figure~\ref{cnoc2mr:figmbz}, $M_B^{k,e}$ is 
plotted versus redshift for all CNOC2 galaxies 
with $R_C \leq 21.5$ and $0.12 \leq z \leq 0.55$.  
We will continue to refer to this figure as we impose more 
restrictions on this initial sample.  

\subsection{Adjusting the Flux Limit For $k$-corrections}\label{cnoc2mr:limits}

\begin{figure}
\plotone{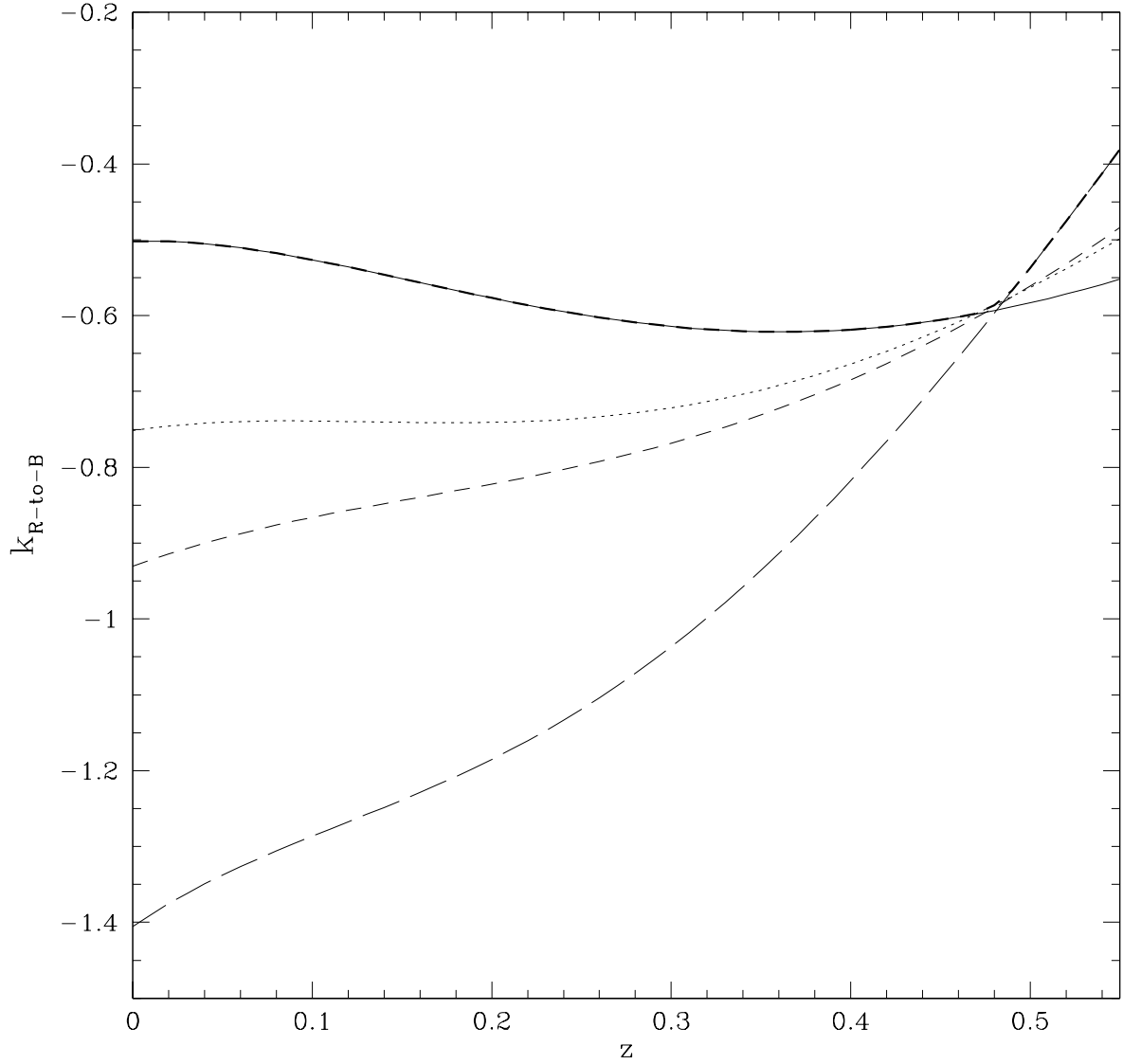}
\caption[Model $k$-corrections]
{
Model $k$-corrections from Coleman, Wu, and Weedman 
(1980) are given for 4 galaxy types.  
Lines are as follows : E/S0 (long-dashed line), Sbc (dashed line), 
Scd (dotted line), and Im (solid line).  
The maximal $k$-correction $\kmax (z)$ is marked with 
a thick dashed line.  
\label{cnoc2mr:figkRtoB}}
\end{figure}

The chosen flux limit imposes a redshift-dependent 
limiting absolute magnitude $\mlim (z)$ on the sample, such that
\begin{equation}\label{eqn:fluxlimit} 
\mlim (z) = \rclim - 5\log d_L(z) - 25 - \krtob (z) - E(z),
\end{equation}
where $\krtob (z)$ gives the $k$-correction and transformation from 
$R_c$ to $B$, and $E(z)$=$-Qz$.
As $k$-corrections vary for galaxies with different SED's, we run the 
risk of preferentially selecting galaxies of a particular 
spectral type.  This applies to galaxies that lie close to 
the flux limit.  To avoid this bias, we choose a more 
conservative limit, to ensure that galaxies of all 
spectral types will be observable.  This is done by 
comparing the $k$-corrections of \citet{CWW} 
for 4 different SED types (E/S0, Sbc, Scd, and Im).  
At each redshift, we select the SED type which, 
for a given $R_C$ apparent magnitude limit, yields the most 
conservative (ie., brightest) $B$-band absolute magnitude limit.
This is shown in Figure~\ref{cnoc2mr:figkRtoB}.  
At $z \lesssim 0.47$, the Im SED is chosen, 
while the E/S0 SED takes over at higher redshift 
($k$-corrections are nearly identical for all galaxy types 
at $z \sim 0.48$).  
This gives us a function, hereafter denoted $\kmax (z)$, 
which provides a good estimate of the maximal $k$-correction 
at all redshifts of interest. 
We now combine this function with the chosen flux limit 
($\rclim$=21.5) to set the limiting absolute magnitude 
as a function of redshift.  
This relation is shown in Figure~\ref{cnoc2mr:figmbz}.  
This constraint ensures that galaxies of all spectral types 
will have an equal probability of falling within our sample.  

\subsection{Minimizing the Luminosity Dependence of Clustering}
\label{cnoc2mr:clust}
When computing pair statistics, it is safest to use a volume-limited 
sample.  However, if we were to impose such a restriction on the CNOC2 survey, 
we would not retain a statistically useful sample of close galaxy pairs.  
Instead, in order to maximize the size of the observed sample, 
we choose to use a flux-limited sample, and correct the pair statistics 
to that of a volume-limited sample.  
As discussed by P2000\nocite{ssrs2mr}, 
it is prudent to constrain a flux-limited 
sample to a relatively small range in absolute magnitude, in order to 
minimize the bias that may be introduced by luminosity-dependent 
clustering.  
In order to address this issue, we further restrict the observed 
sample to the range $-21 \leq M_B^{k,e} \leq -17$.  
The bright limit reduces the sample size by only 0.2\%, and 
reduces concerns about biases due to increased clustering of 
very bright galaxies.  
The faint limit eliminates low luminosity galaxies at 
$z \lesssim 0.23$, reducing the sample size by 3.0\%.  
These constraints are shown in Figure~\ref{cnoc2mr:figmbz}. 
Our final sample consists of 4184 galaxies which satisfy
all of the criteria outlined in this section.  In order to 
maximize the number of pairs observed, this sample will be 
used for choosing both primary (host) and secondary (companion) 
galaxies (see \S~\ref{cnoc2mr:newstats}).

\subsection{Choosing Volume Limits}\label{cnoc2mr:m1m2}

Having outlined the criteria to be used for {\it selecting} galaxies 
from the flux-limited CNOC2 survey, we must now match the sample 
to a suitable range in absolute magnitude, in order to correct the 
pair statistics to those of a volume-limited sample.  
At the bright end, we have restricted the observed sample 
to $M_B^{k,e} \geq -21$ at all redshifts, so this provides a 
natural bright limit for the volume-limited sample.
At the faint end, 
the limiting absolute magnitude of the observed sample varies with redshift, 
from $M_B^{k,e}$=$-17$ at low redshift to $M_B^{k,e}\sim-19$ at $z=0.55$ (see 
Figure~\ref{cnoc2mr:figmbz}).  
The mean limiting absolute magnitude of the sample, computed using weights 
described in the following section, is $M_B^{k,e}$=$-17.9$.  
For convenience, we select a faint limit for the volume-limited 
sample of $M_B^{k,e}$=$-18$.
That is, pair statistics from the observed 
sample (\S~\ref{cnoc2mr:flux}$-$\S~\ref{cnoc2mr:clust}) 
will be normalized to the absolute magnitude range 
$-21 \leq M_B^{k,e} \leq -18$.  
This is identical to the limits chosen by 
P2000\nocite{ssrs2mr} for the SSRS2 sample,  
enabling us to make a direct comparison between our CNOC2 
pair statistics ($z \sim 0.3$) and the SSRS2 pair statistics 
($z \sim 0$).  

\section{ACCOUNTING FOR SELECTION EFFECTS} \label{cnoc2mr:select}

P2000\nocite{ssrs2mr} devised a simple weighting scheme to 
apply when measuring pair statistics for a flux-limited 
redshift survey.  
We will generalize this approach to account for 
several additional selection effects present in CNOC2.  
There are two key points to consider.  First, 
companions should be weighted so as to 
normalize correctly
the number and luminosity of companions to that expected for a volume-limited 
sample with a fixed range in absolute magnitude.  

Secondly, if the intrinsic pair statistics are assumed to be
similar for all galaxies in the sample, one can then apply weights 
to galaxies in the primary sample (so-called host galaxies) 
so as to give larger weights to galaxies that, 
on average, would be {\it expected} to have greater numbers of 
detected companions.  An optimal weighting scheme for the 
primary sample will minimize the measurement error in the pair statistics.  
For example, galaxies at the low redshift 
end of a flux-limited sample would be expected to have more 
detected companions (on average) than galaxies at the 
higher redshift end of the same sample.  It is worth stressing that 
we {\it do not} weight galaxies based on their {\it observed} number of 
companions.

In the following sections, we will apply this methodology 
to several selection effects that are present in our sample.  
We will treat each selection effect separately at first, 
showing how they relate to the weights of galaxies in 
the primary ($w_1$) and secondary ($w_2$) samples.  
Where necessary, we will distinguish between weights 
applicable specifically to $N_c$ ($w_{N_1}$, $w_{N_2}$) and 
$L_c$ ($w_{L_1}$, $w_{L_2}$).
In 
Section~\ref{overall}, we summarize and 
combine these weights to give
expressions for the final weights used in the pairs analysis.  

\subsection{Accounting for the Flux Limit}

We must first determine $\mlim(z)$, which gives the 
limiting absolute magnitude allowed at redshift $z$.  At most 
redshifts, this is given by equation~\ref{eqn:fluxlimit}, 
using the maximal $k$-correction outlined in Section~\ref{cnoc2mr:limits}..
At the low redshift end of the sample, however, 
the faint {\it absolute} magnitude limit imposed 
($M_B^{k,e}$=$-17$; \S~\ref{cnoc2mr:clust}) will take over.  Therefore, 
the limiting absolute magnitude used for identifying galaxies 
in the secondary sample is given by
\begin{equation}\label{cnoc2mr:eqnmlim}
\mlim (z) = \min [-17, \rclim -5\log d_L(z) - 25- \kmax (z) - E(z)].
\end{equation}
At the bright end, the sample is limited by $M_B^{k,e}$=$-21$.   
When relating a flux-limited sample to its volume-limited 
counterpart, it is possible to transform between the two, 
provided the LF is known.  
The selection function, 
denoted $S(z)$, is defined as the density of galaxies expected 
in the flux-limited sample, divided by the density of galaxies 
expected in the volume-limited sample.  
We will use this function to derive appropriate weights 
for our pair statistics.  
The form of the selection function, 
given for both number density ($S_N(z)$) and luminosity 
density ($S_L(z)$), is as follows : 
\begin{equation} \label{cnoc2mr:eqwN2Q}
S_N(z) = 
{\int_{-21-Qz}^{\mlim(z)-Qz} \phi(M,z)dM
\over 
\int_{-21-Qz}^{-18-Qz} \phi(M,z)dM}, 
\end{equation}
\begin{equation} \label{cnoc2mr:eqwL2Q}
S_L(z) = 
{\int_{-21-Qz}^{\mlim(z)-Qz} \phi(M,z)LdM
\over 
\int_{-21-Qz}^{-18-Qz} \phi(M,z)LdM}. 
\end{equation}

To provide an intuitive feel for the effects of the modelled luminosity 
evolution ($Q$) on this correction, we give 
a simple example.  Suppose we 
compute pair statistics for $-21 \leq M_B \leq -18$ at $z$=0, and wish to 
make a direct comparison with a sample at $z$=0.3.  
Assuming $Qz$=0.3 magnitudes of luminosity evolution, we should 
normalize to $-21.3 \leq M_B \leq -18.3$ at $z$=0.3.  
This will increase $S_N(z)$ and $S_L(z)$, and will 
translate into a decrease in $N_c$ and $L_c$.  
In Section~\ref{cnoc2mr:Qdepend}, we will discuss the impact of this 
correction on the main results of this study. 

In order to recover pair statistics that are applicable to a 
secondary sample with $-21 \leq M_B^{k,e} \leq -18$, one should apply weights 
$w_{N_2}(z)\propto 1/S_N(z)$ and 
$w_{L_2}(z) \propto 1/S_L(z)$ to 
all companions at redshift $z$.  
Monte Carlo simulations confirm that 
errors in the pair statistics are minimized by applying 
weights to galaxies in the primary sample that are the 
reciprocal of the secondary weights (P2000\nocite{ssrs2mr}).  
Thus, optimal weighting is given by $w_{N_1}(z)\propto S_N(z)$ and 
$w_{L_1}(z) \propto S_L(z)$.

\subsection{Overall Spectroscopic Completeness}\label{wmc}
The CNOC2 survey, like many other redshift surveys, is not 
spectroscopically complete.  
This incompleteness must be taken into account when computing 
pair statistics.  
CNOC2 selection weights have been computed to account for dependence 
on apparent magnitude ($w_m$), color ($w_c$), 
location ($w_{xy}$), and redshift ($w_z$) \citep{YEC,CNOC2data}.
For this study, we combine these weights 
to arrive at an overall spectroscopic 
weight ($w_s$) for each galaxy, where $w_s$=$w_mw_cw_{xy}w_z$.

First, we apply these weights to 
each close companion, in order to compensate for the 
underestimate in $N_c$ and $L_c$.  Thus, $w_2 \propto w_s$.  
When applying weights to primary galaxies, the idea is to 
give increased weight to galaxies which are likely to have 
larger numbers of detected companions.  In this case, the 
spectroscopic weight of a primary galaxy does not tell 
us whether we are more or less likely to find observed 
companions.  This is because there is no direct correlation 
between the spectroscopic weights of two galaxies in a close 
pair (for example, the two members may have widely differing 
apparent magnitudes).  Therefore, we choose not to apply 
these spectroscopic weights to galaxies in the primary sample.  

We note here that, if spectroscopic weights were correlated 
with the intrinsic pair statistics in the sample, we would 
be required to assign $w_1 \propto w_s$.  This scenario was 
ruled out by computing pair statistics for both choices of 
weighting schemes; the resulting pair statistics were found to 
be very insensitive to the choice of spectroscopic 
weighting scheme used for the primary galaxies.  On the other 
hand, the spectroscopic weighting of the secondary sample is 
essential, and must be applied as described above.

\subsection{Spectroscopic Completeness at Small Separations}
\label{wtheta}

We have accounted for the overall spectroscopic incompleteness 
as a function of apparent magnitude, color, location, and redshift.  
We now investigate if there is any dependence on pair separation.  
Recall that we use two masks per field when acquiring 
spectra (see \S~\ref{cnoc2mr:spectra}).  
While this increases the overall 
completeness of the survey, it also allows for a better 
handling of objects in close pairs.  If two objects are close 
together on the sky, it is usually not possible to place 
slits on both objects simultaneously.  
Hence, if a single mask is used, these 
objects will be systematically underselected.  Our mask design 
program compensates for this effect by giving 
preference to these objects on the second mask \citep{YEC}.  
However, even with two masks, it is not possible to obtain a 
fair sample of close triples and higher order systems.  
This effect is somewhat compensated for using the geometric 
weights ($w_{xy}$), which are smoothed on a scale of 2\arcmin.  
We now check to see how well the mask design algorithm 
has worked, and attempt to measure and correct for any bias 
that remains after the geometric weights ($w_{xy}$) have been applied.

We begin by identifying two samples of galaxies.  The first 
contains all CNOC2 galaxies with $R_C \leq 21.5$, and will 
be referred to as the photometric sample.  We then identify 
a spectroscopic sample, which consists of all galaxies in 
the photometric sample with measured redshifts in the range 
$0.12 \leq z \leq 0.55$.  We will use the ratio of galaxies 
in these two samples to measure spectroscopic completeness.

\begin{figure}
\plotone{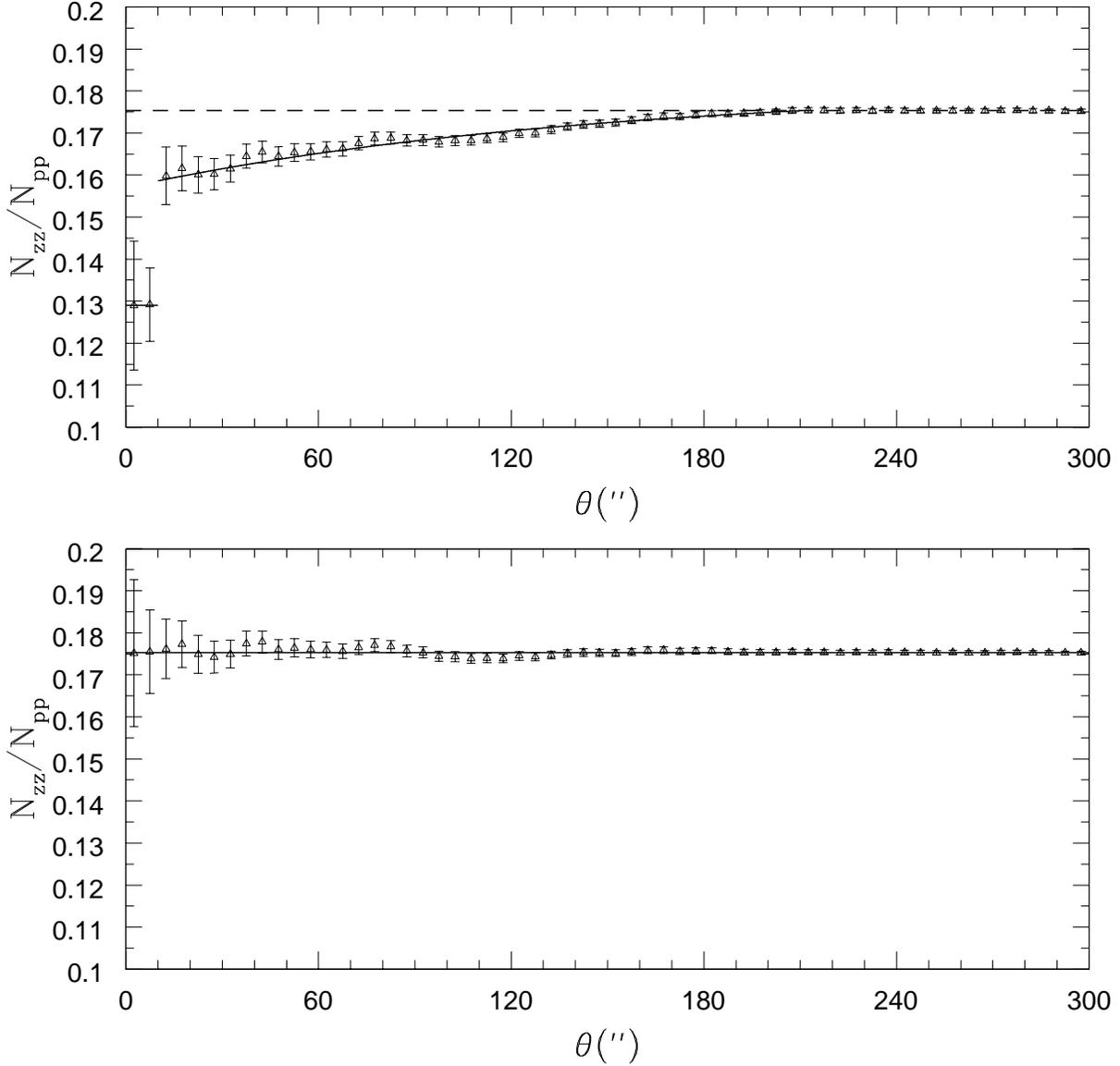}
\caption[Spectroscopic Completeness in Close Pairs]
{Spectroscopic completeness is computed for a range of 
angular pair separations.  
(a) In the upper panel, 
we compute the ratio of spectroscopic pairs 
($N_{\rm zz}$) to photometric  
pairs ($N_{\rm pp}$), as a function of angular separation $\theta$.  
Error bars are computed using the Jackknife technique.  
With fair selection, 
$N_{\rm zz}/N_{\rm pp}=f_s^2 \approx 0.1753$ (dashed line), as this 
gives an excellent fit to the completeness on large scales 
($\theta > 3.5\arcmin$).    
The incompleteness is modelled with a power law at 
$10 \arcsec < \theta < 210\arcsec$.  
This power law clearly does not provide 
an acceptable fit on smaller scales; hence, 
we take $N_{\rm zz}/N_{\rm pp}$=0.129 for 
$\theta \leq 10\arcsec$.  
The combined fit is marked with a solid line.  
(b) The observed spectroscopic completeness 
is corrected using weights from the modelled fit, and 
is plotted in the lower panel.  The horizontal solid line 
gives $f_s^2$=0.1753.  The corrected data points are consistent 
with this value at all separations.
\label{cnoc2mr:complete}}
\end{figure}

We wish to determine how the spectroscopic completeness 
varies as a function of angular pair separation $\theta$.  
We begin by measuring $\theta$ for 
all pairs in the photometric sample.
These pairs will be referred to as p-p pairs.  Similarly, 
we find all z-z pairs in the spectroscopic sample.  
We assign these pairs to 
bins of angular size 5\arcsec, for separations less than
5\arcmin.  Each z-z pair is weighted by the product of the 
geometric weights ($w_{xy}$), since these weights compensate 
in part for the effect of interest here.  
Pairs in the p-p sample are given equal weights, since the 
photometric sample is complete.  
If paired galaxies are selected fairly, the weighted number of 
z-z pairs (hereafter $N_{\rm zz}$) is related to 
the number of p-p pairs ($N_{\rm pp}$) by  
\begin{equation}\label{cnoc2mr:eqnzz}
N_{\rm zz}(\theta)=f_s^2N_{\rm pp}(\theta), 
\end{equation}
where $f_s$ is the mean spectroscopic completeness on 
large scales.  
We compute $N_{\rm zz}(\theta)/N_{\rm pp}(\theta)$ for 
the full CNOC2 sample, and compute error bars 
using the Jackknife technique.  
The results are given in the upper panel of 
Figure~\ref{cnoc2mr:complete}.  

It is immediately apparent that there is a significant and 
systematic decrease in spectroscopic completeness at small 
separations.  
This deficit becomes noticeable at 3.5\arcmin, 
and increases fairly smoothly down to $\approx 10\arcsec$.  
A sharp drop is seen below $10\arcsec$.
These results clearly indicate that our mask 
design algorithm and geometric weights do not 
completely compensate for pair 
selection effects.  Without correction, this would lead 
to a very significant underestimate in our pair statistics 
(note that most of the close pairs used in this study have 
$\theta \leq 10\arcsec$).  

We correct for this effect by modelling the incompleteness, 
such that
\begin{equation}\label{cnoc2mr:eqngth}
N_{\rm zz}(\theta)=g(\theta)N_{\rm pp}(\theta).
\end{equation} 
On large scales, $g(\theta)$ is independent of pair separation, 
as expected.  On intermediate scales, $g(\theta)$ can be fit 
with an exponential function.  This trend does not continue 
to the smallest scales ($\theta < 10\arcsec$);
hence, we take $g(\theta)$ to 
be a constant in this regime.  The resulting model fit 
for $g(\theta)$ is given by 

\begin{eqnarray}
g(\theta)=\left\{
\begin{array}{cc}
0.129;& \theta \leq 10\arcsec\\
-0.024{\rm e}^{-0.007\theta}+0.181;&
10\arcsec < \theta \leq 210\arcsec\\
0.175;& \theta > 210\arcsec.\\
\end{array}\right\}
\end{eqnarray}

This function gives a good match to the data at all angular 
separations, and is shown in the upper panel of 
Figure~\ref{cnoc2mr:complete}.  

Using this functional fit, we are able to 
estimate the deficit in spectroscopic completeness as a 
function of angular separation.  To remove this deficit, 
we assign each companion a weight, denoted $w_{\theta}$, 
that is inversely proportional to the deficit.  That is, 
$w_{\theta} = f_s^2/g(\theta)$.
We repeat the measurement 
of spectroscopic completeness using these weights, and 
plot the results in the lower panel of 
Figure~\ref{cnoc2mr:complete}.  The corrected measurements 
of $N_{\rm zz}/N_{\rm pp}$ are consistent at all separations 
less than 5$\arcmin$, within the errors.  
Thus, this weighting scheme successfully removes the bias 
due to decreased spectroscopic completeness on small scales.  

We must now incorporate these weights ($w_{\theta}$) into the 
measurement of pair statistics.  The first task is to ensure 
that we apply weights to the secondary sample such that the 
correct number or luminosity of companions will be recovered.  
Clearly, for each companion at separation $\theta$, one should 
apply weight $w_{{\theta}_2} = w_{\theta}$.  
For the close companions found in this study (\S~\ref{cnoc2mr:nclc}), 
the mean angular separation is $5\arcsecpt 0$, yielding 
a mean weight of $w_{{\theta}_2}$=$1.36$.  
The net effect of 
these weights is to increase $N_c$ and $L_c$ 
by $\approx$ 35\%.  
We do not apply $w_{\theta}$ weights to galaxies 
in the primary sample, as these weights are relevant 
only for galaxies with close companions.

\subsection{Boundary Effects}
Some of the galaxies in the primary sample lie close to 
the edge of the field (on the sky), or within $\Delta \vmax$ 
of the redshift limits.  In addition, 
a number of galaxies lie close to 
bright stars; consequently, some of the surrounding regions 
may be hidden from view.
Each of these factors will contribute 
to an underestimate of the pair statistics.  

For CNOC2, we have 
generated field area maps which mark out the edges of 
each patch, and indicate which regions are blocked by bright 
stars.  For each galaxy in the primary sample, 
we compute the fraction of the sky 
within $\rpmin \leq r_p \leq \rpmax$ that lies within these survey 
boundaries, where $\rpmin$ and $\rpmax$ 
denoted the minimum and maximum projected separation 
used to define close companions.  
This fraction will be denoted $\fb$.  
For CNOC2, our usual choices of $\rpmin$ and $\rpmax$ (see 
\S~\ref{cnoc2mr:methods}) lead to $\fb$ = 1 for 94.7\% 
of the galaxies in the primary sample.  
For the remainder, most have $\fb$ 
close to 1, with a total of only 0.2\% having $\fb < 0.5$.  
Each companion is assigned a boundary weight $w_{b_2} = 1/\fb$, 
where $\fb$ is associated with 
its host galaxy from the primary sample.  
By multiplying each companion by 
its boundary weight, we will recover the correct number of 
companions.  To minimize errors, primary galaxies are 
assigned weights $w_{b_1}$=$\fb$.  

We now consider galaxies which lie near the survey 
boundaries along the line of sight.  
We follow the approach laid out by P2000\nocite{ssrs2mr}.  
That is, we exclude  
all companions that lie between a primary galaxy and its nearest 
redshift boundary, provided the boundary lies within 
$\Delta \vmax$ of the primary galaxy.  
We assign a weight of $w_{2_v}$=2 to any companions found 
in the direction opposite to the boundary.  
To minimize the errors in computing the pair statistics, 
the corresponding primary galaxies are 
assigned weights $w_{v_1}$=0.5.  

\subsection{Combining Weights}\label{overall}
To summarize, for a primary galaxy at redshift $z_i$, 
weights for companions in the secondary sample 
are given by :
\begin{equation}\label{cnoc2mr:eqnwn2}
w_{N_2}=S_N(z_i)^{-1} w_{s_2}w_{{\theta}_2}w_{b_2}w_{v_2},
\end{equation}
\begin{equation}\label{cnoc2mr:eqnwl2}
w_{L_2}=S_L(z_i)^{-1} w_{s_2}w_{{\theta}_2}w_{b_2}w_{v_2},
\end{equation}
where $S_N(z_i)$ and $S_L(z_i)$ are given by 
equations~\ref{cnoc2mr:eqwN2Q} and~\ref{cnoc2mr:eqwL2Q}
respectively.
For the primary sample, the corresponding expressions are 
as follows : 
\begin{equation}\label{cnoc2mr:eqnwn1}
w_{N_1}=S_N(z_i) w_{b_1}w_{v_1},
\end{equation}
\begin{equation}\label{cnoc2mr:eqnwl1}
w_{L_1}=S_L(z_i) w_{b_1}w_{v_1}.
\end{equation}
The total number and luminosity of close companions for the 
$i^{th}$ primary galaxy, computed by summing over the $j$ 
galaxies satisfying the close companion criteria, 
is given by 
$N_{c_i} = \sum_j{w_{N2}(z_j)}$ and
$L_{c_i} = \sum_j{w_{L2}(z_j)L_j}$ respectively.
The mean number and luminosity of close companions is then 
computed by summing over all galaxies in the primary sample, 
using weights $w_{N1}$ and $w_{L1}$, yielding 
\begin{equation} \label{cnoc2mr:eqNcobs}
{N_c} = {\sum_i w_{N1}(z_i)N_{c_i} \over \sum_i w_{N1}(z_i)}
\end{equation}
\begin{equation} \label{cnoc2mr:eqLcobs}
{L_c} = {\sum_i w_{L1}(z_i)L_{c_i} \over \sum_i w_{L1}(z_i)} .
\end{equation}

\section{CNOC2 PAIR STATISTICS} \label{cnoc2mr:nclc}
\begin{figure}
\plotone{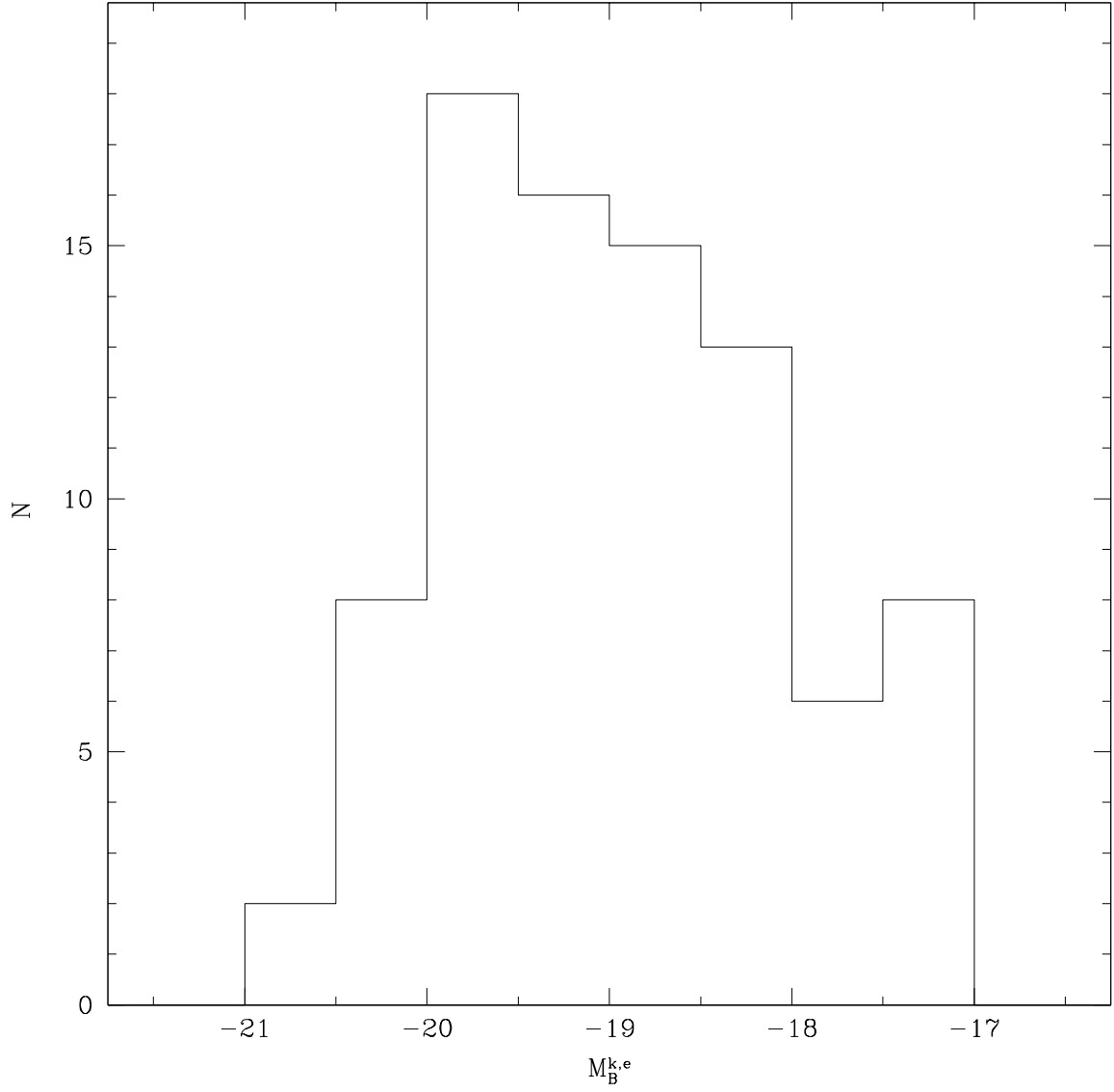}
\caption[Absolute Magnitude Histogram for Companions]
{An absolute magnitude histogram is given for the 88 companions 
used in the pair statistics.  
\label{cnoc2mr:figlh}}
\end{figure}

We have now set out an approach for measuring pair 
statistics for the CNOC2 survey.  
We define a close companion to be one with a projected 
physical separation of 5~\hkpc~$\leq r_p \leq$~20~\hkpc~ 
and a rest-frame line-of-sight velocity difference of 
$\Delta v \leq$ 500 km/s.  
Using this definition, and the survey parameters set out 
above, we find a total of 88 close companions in CNOC2.  
When using the same range in absolute magnitude for the 
primary and secondary samples ($-21 \leq M_B^{k,e} \leq -18$) 
as we have done here, a given 
galaxy pair usually contributes two companions.  For CNOC2, 
this is the case for all of our pairs; furthermore, no 
spectroscopic triples are found.  Thus, our 88 close 
companions are found in 44 unique pairs.  
In Tables~\ref{cnoc2mr:tabcp02}$-$\ref{cnoc2mr:tabcp21}, 
we present the following properties of these pairs: 
Pair ID, CNOC2 ID (PPP number) for each galaxy, projected 
physical separation ($r_p$) and rest-frame line-of-sight velocity 
difference ($\Delta v$) of the galaxies, coordinates of the pair 
center, and the mean redshift of the pair.
A histogram of companion absolute magnitudes is given 
in Figure~\ref{cnoc2mr:figlh}.
We also present a mosaic of $R_C$ images for these 
systems in Figure~\ref{cnoc2mr:figim}.  Some of 
these pairs exhibit clear signs of interactions; however, 
in most cases, the poor resolution of these ground based 
images renders classification uncertain at best.  
Hubble Space Telescope imaging of these pairs will be presented 
in a forthcoming paper (D.R. Patton, in preparation).  

Using this sample of companions, pair statistics 
were computed for each of the 4 CNOC2 patches.  
Errors were computed using the Jackknife technique.  
These results are given in Table~\ref{cnoc2mr:tabstats}, 
along with the number of galaxies used in each sample (N) 
and the observed number of companions (${\rm N}_{{\rm comp}}$).  
Results from the 4 patches were combined, 
weighting by Jackknife errors, to give 
$N_c = 0.0321 \pm 0.0077$ 
and $L_c = 0.0294\pm 0.0084 \times 10^{10}~h^2 \lsun$ 
at $z$=0.30.  
Results from all 4 patches are consistent with these 
mean values, within the quoted 1$\sigma$ errors. 
We now investigate how sensitive these results are to 
the particular parameters selected in this study.  

\subsection{Dependence on Limiting Absolute Magnitude}\label{cnoc2mr:dependm2}
We have stressed the importance of specifying a range in 
absolute magnitude for companions, when computing 
pair statistics.  The faint limit (hereafter $M_2$) is particularly 
important.  For this study, we have selected 
$M_2(B)$=$-18$ as being representative of our sample 
(see~\S~\ref{cnoc2mr:m1m2}).  
It is useful to see how our results change with different 
choices of this important parameter.  
We now compute our pair 
statistics for $-19 \leq M_2 \leq -17$.  
The results, after combining all 4 patches, are given 
in Table~\ref{cnoc2mr:tabm2}.  

Both $N_c$ and $L_c$ are expected to increase as $M_2$ 
becomes fainter, and this is seen in Table~\ref{cnoc2mr:tabm2}.  
$N_c$ increases by a factor of $\sim 5$ between 
$M_2$=$-19$ and $M_2$=$-17$.  $L_c$ is less sensitive to 
$M_2$, changing by a factor of $\sim 2$ over the same 
range.  In both cases, the changes are due solely to 
the increase in mean number or luminosity density, resulting 
from integrating deeper into the LF.  These trends are 
very similar to what P2000\nocite{ssrs2mr} 
found for SSRS2, and further emphasize the importance of 
accounting for sample depth when computing pair statistics.  

\subsection{Dependence on $\rpmax$}
Close companions are required to have projected separations 
less than $\rpmax=20$~\hkpc.  
While this maximum separation is thought to be ideal for 
isolating good merger candidates, it is useful to see 
how the pair statistics behave at larger separations.  
With this in mind, we compute pair statistics for 
10 \hkpc $\leq \rpmax \leq 100~$\hkpc, 
with $\Delta \vmax = 500$ km/s.  
Results are given in Figure~\ref{cnoc2mr:figrp}.  
This plot indicates a smooth increase in both $N_c$ and $L_c$  
with $\rpmax$.  This trend is expected from measurements 
of the galaxy correlation function.  
This function is commonly expressed as a power law of the 
form $\xi(r,z) = (r_0/r)^{\gamma}$, 
with $\gamma$=1.8 \citep{DP83}.
Integration over this function yields pair statistics that 
vary as $r_p^{3-\gamma} \approx r_p^{1.2}$, which is in 
good agreement with the trend found in Figure~\ref{cnoc2mr:figrp}.

\subsection{Dependence on $\Delta \vmax$}
We also compute pair statistics for a range in $\Delta \vmax$.  
This is done first for $\rpmax = 20 $~\hkpc, showing the 
relative contributions at different velocities to the mean 
pair statistics quoted in this study.  We also compute 
statistics using $\rpmax = 50~$\hkpc, in order to improve the 
statistics.  
Results are given in Figure~\ref{cnoc2mr:figrl}.  
Two important conclusions may be drawn from this plot.  
First, at small velocities, 
both $N_c$ and $L_c$ increase with $\Delta \vmax$, as 
expected.  This simply indicates that one continues to find 
additional companions as the velocity threshold increases.  
Secondly, the pair statistics begin to flatten out at around 
$\Delta \vmax$=500 km/s.  Increasing the threshold to 
1000 km/s or higher does increase the size of the pair 
sample substantially.  Moreover, the additional pairs that would 
be found would be less likely (on average) to be good merger 
candidates.  We conclude that $\Delta \vmax = 500$ km/s 
is a sensible and efficient choice.  The main results of this 
study are unaffected by small changes in the choice of 
$\Delta \vmax$.

\section{MERGER RATE EVOLUTION} \label{cnoc2mr:mrate}
Having computed pair statistics at moderate redshift, we 
now compare our measurements 
with the results of P2000\nocite{ssrs2mr} 
to infer evolution in the galaxy merger and 
accretion rates from $z \sim 0.3$ to $z \sim 0$.  We begin by 
justifying a direct comparison of these two samples.  
After estimating the evolution in the merger and accretion 
rates, we will explore the sensitivity of our results to 
various assumptions that have been made in this analysis.  

\subsection{Validity of Comparison}
Throughout this study, we have stressed the importance 
of making careful measurements of pair statistics, to 
avoid various biases that may adversely affect the results.  
Here, we review the important issues that must be addressed 
before comparing pair statistics for different samples.  

First of all, one must ensure that the definition of a close 
companion is identical in all samples.  When comparing samples 
at different redshifts, one must be cautious of definitions 
that may have redshift-dependent biases present. 
For example, earlier photometric studies of galaxy pairs 
(e.g., Zepf \& Koo 1989\nocite{ZK}) had to correct for 
optical contamination due to unrelated foreground or 
background galaxies.  
While this contamination can be accounted for, the degree of 
contamination increases systematically with redshift; hence, 
it is clearly preferable to avoid this correction, by using  
companions with measured redshifts.  
For both SSRS2 and CNOC2, we have used the same definition of 
a close companion.  Our dynamical definition is unaffected 
by optical contamination, or other redshift-dependent biases.
The only remaining factor that may cause our definition to change
with redshift is the choice of cosmology.  That is, if our choice 
of cosmological parameters is not correct, there will be a 
redshift-dependent change in the projected physical separation 
used to identify close companions.  
In Section~\ref{cnoc2mr:cosm}, we explore the effects that 
different choices of cosmological parameters have on 
our estimates of merger rate evolution.  

There are two passband effects that must be considered 
when comparing different samples.  First, any limits in 
absolute magnitude must be the same for all samples.  
For both SSRS2 and CNOC2, we compute pair statistics in 
rest-frame $B$.
It is also important to {\it select} galaxies at approximately 
the same rest-frame wavelength.  If this is not done, any 
differences in the resulting pair statistics may simply 
be artifacts of the selection process.
The SSRS2 sample was selected in $B$.  
The CNOC2 sample was selected in observed 
$R_C$, which is roughly equivalent to rest-frame $B$ at 
$z\sim 0.4$.  Thus, SSRS2 and CNOC2 are selected at 
comparable rest-frame wavelengths.  

Finally, we have emphasized the need to specify a range in 
absolute magnitude for computing pair statistics.  
When comparing different samples, it is critical that 
this correspond to the same intrinsic luminosity in all samples.  
P2000\nocite{ssrs2mr} computed pair statistics for SSRS2 
using $-21 \leq M_B \leq -18$.  
For CNOC2, we have chosen the same range in absolute magnitude,   
accounting for $k$-corrections and luminosity evolution.  This helps 
to ensure that we are probing the same range in galaxy masses  
at all redshifts.  

\subsection{Merger Rate Evolution Using $\rpmax$ = 20 \hkpc}
\label{cnoc2mr:nclcz}
Having demonstrated that it is reasonable to compare 
our SSRS2 and CNOC2 pair statistics, we now proceed 
with the analysis.  
P2000\nocite{ssrs2mr} computed pair statistics 
for the SSRS2 survey \citep{SSRS2}, 
which consists of 5426 galaxies at $z\sim 0$.  
Primary and secondary samples were drawn from the same 
set of galaxies.
Using close (5 \hkpc~ $\leq r_p \leq$ 20 \hkpc) 
dynamical ($\Delta v \leq 500$ km/s) pairs, 
the resulting pair statistics were as follows : 
$N_c(-21 \leq M_B^{k,e} \leq -18)=0.0226 \pm 0.0052$ and 
$L_c(-21 \leq M_B^{k,e} \leq -18) = 0.0216 \pm 0.0055 \times 10^{10}~h^2 \lsun$ 
at $<$$z$$>$=0.015.  

CNOC2 pair statistics were given 
in Table~\ref{cnoc2mr:tabstats}. 
Pair statistics for both SSRS2 and CNOC2 are plotted in 
the lower portion of Figure~\ref{cnoc2mr:figmrate1}.  
Recall that $N_c$ is related to the galaxy merger rate, 
while $L_c$ depends on the mass accretion rate.  
Following the convention in this field, 
we choose to parameterize evolution in the galaxy merger 
rate by $(1+z)^{m_N}$, where $m_N$ is determined by 
changes in $N_c$ with redshift.  Similarly, we take the 
accretion rate to evolve as $(1+z)^{m_L}$.  
We find $m_N=1.44 \pm 1.34$ and $m_L=1.34 \pm 1.55$.  
These relations are plotted in Figure~\ref{cnoc2mr:figmrate1}.  

\subsection{Merger Rate Evolution Using $\rpmax$ = 50 \hkpc}
\label{cnoc2mr:nclcz50}
The preceding results provide some support for a mild rise in the 
merger and accretion rates with redshift.  However, the error 
bars are uncomfortably large.  There simply are not enough 
small separation pairs in our sample to give a definitive 
answer.  It is possible to improve on this result by relaxing 
our definition of a close pair.  In particular, 
Figure~\ref{cnoc2mr:figrp} indicates that 
both $N_c$ and $L_c$ vary smoothly with $\rpmax$.  
In order to get a better handle on the evolution in 
the merger and accretion rates, we therefore choose to increase $\rpmax$ 
to 50 \hkpc.  Pair statistics were computed for both CNOC2 and SSRS2, 
and are given in Table~\ref{cnoc2mr:tabstats50} (see also 
Figure~\ref{cnoc2mr:figmrate1}).   
These estimates yield merger and accretion rates which evolve 
as $m_N=2.26 \pm 0.70$ and $m_L=2.28 \pm 0.89$.  We will take these 
to be our most secure estimates of evolution in the merger and 
accretion rates.

\subsection{Dependence on Modelling of Luminosity Evolution}
\label{cnoc2mr:Qdepend}
In Section~\ref{cnoc2mr:babs}, we outlined our approach for modelling 
luminosity evolution in our sample, using the $Q$ parameter derived 
from measurements of the CNOC2 LF \citep{CNOC2LF}.  When computing 
pair statistics, we have taken $Q$=1, which assumes an 
average of $z$ magnitudes of luminosity evolution at redshift $z$.  
To see how this assumption affects our results,  
we recompute the pair statistics (with $\rpmax$=50 \hkpc) 
using different choices of $Q$. 
For $Q$=0 (no evolution) we find $m_N=3.07 \pm 0.72$ and $m_L=3.85 \pm 0.85$.  
Thus, if we do not account for luminosity evolution, 
we infer a stronger increase in the merger and accretion rates 
with redshift. 
For $Q$=2, we find $m_N=1.40 \pm 0.67$ and $m_L=0.70 \pm 0.94$. 
In both cases, the effects are strongest for $L_c$, which has an additional 
luminosity, and hence Q, dependence.  
The size of these effects demonstrates the 
importance of compensating for luminosity evolution when 
computing pair statistics.  Nevertheless, for physically plausible 
choices of $Q$, our estimates of $m_N$ and $m_L$ are not dominated 
by uncertainty in the amount of luminosity evolution that is present.

\subsection{Dependence on Cosmology}
\label{cnoc2mr:cosm}
The choice of cosmological parameters affects the computed value of 
$r_p$ for each pair, and also changes the inferred luminosity 
of all galaxies in the sample.  It is therefore important to 
determine the dependence of $m_N$ and $m_L$ on these parameters.
At $z$=0.3, $r_p$ is $\sim 6\%$ smaller for $q_0$=0.5 than 
for our choice of $q_0$=0.1.  Therefore, for a fixed maximum 
projected separation ($\rpmax$), this increases the angular 
search area, and thereby the number of close companions.  
However, this effect is countered by the decreased luminosities 
that result from reduced luminosity distances (see eq.~\ref{eqn:absmb}).
Decreasing galaxy luminosities lowers the number and luminosity density 
of galaxies in the range of interest here ($-21 \leq M_B^{k,e} \leq -18$).
The choice of $q_0$ also affects measurement of the galaxy 
LF, which is needed for measuring pair statistics for this 
flux-limited sample.  
\citet{CNOC2LF} have measured the 
CNOC2 LF for both $q_0$=0.1 and $q_0$=0.5.   
The SSRS2 and CNOC2 pair statistics were recomputed using $q_0$=0.5 
and $\rpmax$=50 \hkpc.  
We find $m_N= 2.13\pm 0.67$ and $m_L= 1.91\pm 0.88$.  That is, 
while a small reduction is seen in the evolution of $N_c$ and $L_c$, 
the competing effects of decreased $r_p$ and decreased luminosities 
nearly cancel out.

The pair statistics may also change in the presence of a 
non-zero cosmological constant.  Using values of 
$\Omega_M$=0.2 and $\Omega_{\Lambda}$=$0.8$, 
$r_p$ increases by about $\sim 6\%$ at $z$=0.3.  
This effect is comparable in size but opposite in 
direction to the example given in the preceding paragraph.  
This implies that this non-zero $\Lambda$ cosmology would 
causes a small increase in the amount of evolution in the merger 
and accretion rates.
We conclude that our pair statistics are insensitive 
to the choice of $q_0$ and $\Omega_{\Lambda}$.
The choice of cosmological parameters will become more of an issue 
as studies of galaxy pairs are extended to higher redshifts
(e.g., Carlberg et al. 2000a\nocite{CFGRS}).

\section{DISCUSSION}\label{cnoc2mr:discuss}

\subsection{Summary of New Results}
We have used the CNOC2 redshift survey to compute secure 
measurements of close pair 
statistics at $z\sim 0.30$.  These are the first 
measurements at $z>0$ that use only dynamically confirmed 
pairs ($\Delta v < 500$ km/s).  Moreover, we have 
carefully accounted for a number of selection biases
that may have adversely affected earlier estimates. 
In particular, we have accounted 
for the dependence on the limiting absolute magnitude 
of companions; without this crucial step, it is 
very dangerous to compare pair statistics from different 
surveys.  This is also the first study to include an 
explicit correction for the redshift-dependent 
bias introduced by luminosity evolution.  

Following the techniques outlined by P2000\nocite{ssrs2mr}, 
we have computed pair statistics using 
the number and luminosity of companions.  
$N_c$ gives the number of close companions per galaxy, while 
$L_c$ measures the luminosity in close companions, per galaxy.
We find $N_c(-21 \leq M_B^{k,e} \leq -18) = 0.0321 \pm 0.0077$ and 
$L_c (-21 \leq M_B^{k,e} \leq -18) = 0.0294\pm 0.0084 
\times 10^{10}~h^2 \lsun$ at $z$=0.30.  
After increasing the maximum pair separation to 50 \hkpc,
and comparing with the low redshift SSRS2 pairs sample, 
we infer evolution in the galaxy merger and accretion rates
of about $(1+z)^{2.3 \pm 0.7}$ and $(1+z)^{2.3 \pm 0.9}$
respectively.  

\subsection{Comparison With Earlier Studies}
As with earlier close pair studies, we have arrived at 
an estimate of the degree of evolution in the galaxy merger rate.
These studies have yielded a wide variety of results 
in the past (see Patton et al. 1997\nocite{P97} for a review).  
Using a consistent transformation between the pair fraction 
and merger rate, results parameterized by $(1+z)^m$
vary from $m\sim 0$ \citep{WOODS} 
to $m\sim 5$ \citep{ZK,YE}.  
After accounting for various discrepancies due to 
optical contamination and spectroscopic completeness, 
these results were shown to be roughly consistent 
with the value of $m=2.8 \pm 0.9$ derived by \citet{P97}.  
More recently, \citet{CFRS00} 
estimated the merger rate to 
evolve as $m=2.7 \pm 0.6$, using Hubble Space Telescope imaging of 
the CFRS and LDSS redshift surveys.

Despite the apparent convergence in results, 
one must be very careful when comparing samples 
of close pairs at different redshifts.  If differences 
exist in the limiting absolute magnitudes of the samples, 
or if there are systematic differences in the way 
galaxies and pairs are selected, any resulting merger rate 
estimate may be strongly affected by redshift-dependent selection effects 
(P2000\nocite{ssrs2mr}).  In our judgement, 
this paper provides the first published estimate of 
merger rate evolution that satisfies these important criteria.  

It is worth noting that, despite an order of magnitude increase in 
redshift survey sizes, the uncertainties in our merger rate estimates 
are still comparable in size to those of \citet{P97} 
and \citet{CFRS00}. 
There are several reasons for this.  
The primary difference in our pair sample is that we have 
required all pairs to have redshifts for both members.  
This yields results that are on a much more secure footing 
than earlier studies which incorporated the use of 
companions with and without redshifts.  We 
have purified our pair sample further by requiring 
$\Delta v \leq 500$ km/s.  We have also restricted our 
sample in redshift and luminosity, in order to minimize 
the detrimental effects of luminosity-dependent clustering.  
All of these effects have made our sample more secure, 
but have greatly reduced the size of our sample.  
Finally, we have computed uncertainties using the Jackknife 
technique, which is well-suited to the weighted 
pair statistics used in this study.  
These error estimates were found to be larger than Poisson 
statistics, by a factor 
of $\approx \sqrt{2}$.  The reason for this difference 
lies in a subtle but important detail regarding the 
application of Poisson statistics to pairs.  The number 
of {\it independent} objects being counted is the 
number of pairs, rather than the number of galaxies 
in pairs.  In \citet{P97}, as in earlier 
studies, this was not recognized, leading to 
uncertainties that were underestimated by a factor
of $\sqrt{2}$.

\subsection{Implications}
In order to interpret our merger rate estimates, 
it is useful to consider several simple scenarios.  
First, for a universe with fixed co-moving density 
and no clustering, the physical density increases 
as $(1+z)^3$.  As our pair statistics measure the 
number and luminosity of companions within a fixed 
physical volume (determined by $\rpmax$ and $\Delta \vmax$), 
they would be expected to increase at 
the same rate.  Of course, there is likely to be 
evolution in the co-moving number density of galaxies, 
as indicated from studies of the galaxy LF.  These changes 
will translate, on average, into comparable changes in 
pair statistics.  
We must also consider the effects of clustering on the 
evolution of the merger rate.  
The scenario outlined above includes 
no clustering, and hence is not representative 
of the real universe.  
In order to incorporate clustering, 
we refer to the galaxy correlation function, which 
provides a convenient parameterization of clustering.  
This function, and its evolution with 
redshift, is traditionally parameterized as follows : 
\begin{equation}\label{cnoc2mr:eqncfz}
\xi(r,z)=\xi(r,0)(1+z)^{-(3+\epsilon)},
\end{equation}
where 
\begin{equation}\label{cnoc2mr:eqnxi0}
\xi(r,0)=\left ({r \over r_0}\right )^{-\gamma}.
\end{equation}
Our pair statistics are proportional to the mean physical 
density, which (in the absence of LF evolution) 
varies as $(1+z)^3$, multiplied by 
an integral over the correlation function, 
which varies as $(1+z)^{-(3+\epsilon)}$.  
Thus, $N_c \propto (1+z)^{-\epsilon}$ ($L_c$ has the same 
dependence).  Suppose we require that 
clustering remain fixed in proper (physical) coordinates.  
This corresponds to $\epsilon$ =0.  
In this case, the physical density of companions will not 
evolve.  That is, we would expect to find $m_N$=$m_L$=0.  
Similarly, we consider a scenario in which the clustering 
is fixed in co-moving coordinates.  In this case, 
$\epsilon=\gamma-3$.  Measurements of the correlation function 
routinely give $\gamma$ = 1.8 \citep{DP83,CNOC2CF}.  
This gives pair statistics that vary as $(1+z)^{1.2}$.  Finally, 
we consider measurements of clustering evolution.  
\citet{CNOC2CF} find $\epsilon=-0.17\pm 0.18$, 
using the CNOC2 survey.  These results are measured on scales of roughly 
0.1 to 10 \hmpc.  If we extrapolate their correlation function 
measurements down to the small scales of interest 
here, we would expect clustering evolution to cause our 
pair statistics to vary roughly as $(1+z)^{0.17}$.  
Our results are inconsistent with this simple clustering 
evolution scenario at the 3$\sigma$ level.  

\subsection{The Cumulative Effect of Mergers Since $z \sim$ 1}
Following the merger remnant analysis of P2000\nocite{ssrs2mr}, 
we will attempt to interpret the implications of 
our results for galaxies at the present epoch.  
We will estimate the fraction of present day galaxies 
that have undergone mergers in the past, referring 
to this quantity as the remnant fraction ($\frem$).  
Suppose that a fraction $\fmg$ of galaxies are undergoing 
mergers at the present epoch.  
If we consider a lookback time of $N\tmg$, 
where $\tmg$ is the merger timescale and $N$ is an integer, 
then 
\begin{equation}\label{ssrs2mr:eqnremz}
\frem = 1 - \prod_{j=1}^N{1- \fmg(z_j) \over 1 - 0.5 \fmg(z_j)},
\end{equation}
where $z_j$ corresponds to a lookback time of $t=j\tmg$. 
We begin by using the estimates of P2000\nocite{ssrs2mr} for the 
local epoch merger fraction ($\fmg$=0.011) 
and the merger timescale ($\tmg$=0.5 Gyr).   
We now employ the estimates of merger rate evolution from 
the present study.  
We take the merger rate to evolve as 
$(1+z)^{2.3}$.  Using equation~\ref{ssrs2mr:eqnremz}, 
with lookback time computed using $h$=0.7 
and $q_0$=0.5, we find that 15\% of $-21 \leq M_B \leq -18$ galaxies 
at the present epoch have undergone a major merger since 
$z\sim 1$.  This remnant fraction is roughly twice as large as
the no-evolution remnant fraction (P2000\nocite{ssrs2mr}).
Even so, our result implies that the majority of bright 
galaxies have not undergone a major merger since $z \sim 1$.  
We note also that our estimated remnant fraction is similar 
to the fraction of bright field galaxies that are ellipticals 
(e.g., Dressler 1980; Postman \& Geller 1984\nocite{MD,PG84}).  
This is consistent with a scenario in which the mergers of 
bright galaxies produce elliptical galaxies.

\subsection{Future Work}

It is now clear that only a few percent of bright galaxies are found 
in close pairs, at least at the modest redshifts under consideration.  
This makes it challenging to find enough dynamical pairs for 
statistically useful measurements of pair statistics, 
even with redshifts surveys of about 5000 galaxies.  
It is even more difficult for sparsely sampled surveys, since the 
observed number of pairs scales with the square of the 
spectroscopic completeness.  
Further progress may require a relaxation of  
the requirement that both members of each pair have measured redshifts.  
We intend to apply our pair statistics to deep imaging of faint galaxies 
in the vicinity of the CNOC2 galaxies used in this survey.  This approach 
will have the added benefit of extending pair statistics to 
the regime of minor mergers. 

Redshift surveys that probe out to higher redshifts will provide more 
leverage on evolution in the galaxy merger and accretion rates (e.g., 
Carlberg et al. 2000a\nocite{CFGRS}).
Our observations - and most models of galaxy formation - indicate that merging 
is likely to be more prevalent at higher redshifts.
However, these samples will be even more susceptible to redshift-dependent 
selection effects (e.g., $k$-corrections and luminosity evolution); 
as a result, it will be critical that the pair statistics be corrected 
for these biases.  
High redshift surveys used to detect companions based on photometry alone 
will face the additional challenge of accounting for increased numbers 
of foreground galaxies, 

Finally, any sample of close galaxy pairs can also be used to study the 
effects of interactions on the galaxies themselves.  
Of particular importance is the question of enhanced star formation 
that may result from these close encounters.  
We have undertaken a number of detailed studies of 
the CNOC2 galaxy pairs identified in this paper, including high resolution 
optical imaging with the Hubble Space Telescope 
(D.R. Patton, in preparation), sub-millimeter continuum 
observations with the James Clerk Maxwell Telescope, 
and multi-color imaging and 
spectroscopy from the CNOC2 survey itself.

\acknowledgments
This work was supported in part by the 
Natural Sciences and Engineering Research Council 
of Canada (NSERC), through a Collaborative Program grant, 
and research grants to C.J.P., H.K.C.Y., and R.G.C..  
H.L. acknowledges support provided by NASA through Hubble Fellowship grant
\#HF-01110.01-98A awarded by the Space Telescope Science Institute, which 
is operated by the Association of Universities for Research in Astronomy, 
Inc., for NASA under contract NAS 5-26555.
Finally, we wish to thank CTAC of CFHT for their generous 
allotment of telescope time, and we gratefully acknowledge the 
support of the CFHT staff.

\clearpage

%% Use the figure environment and \plotone or \plottwo to include 
%% figures and captions in your electronic submission.

\begin{figure}
\plotone{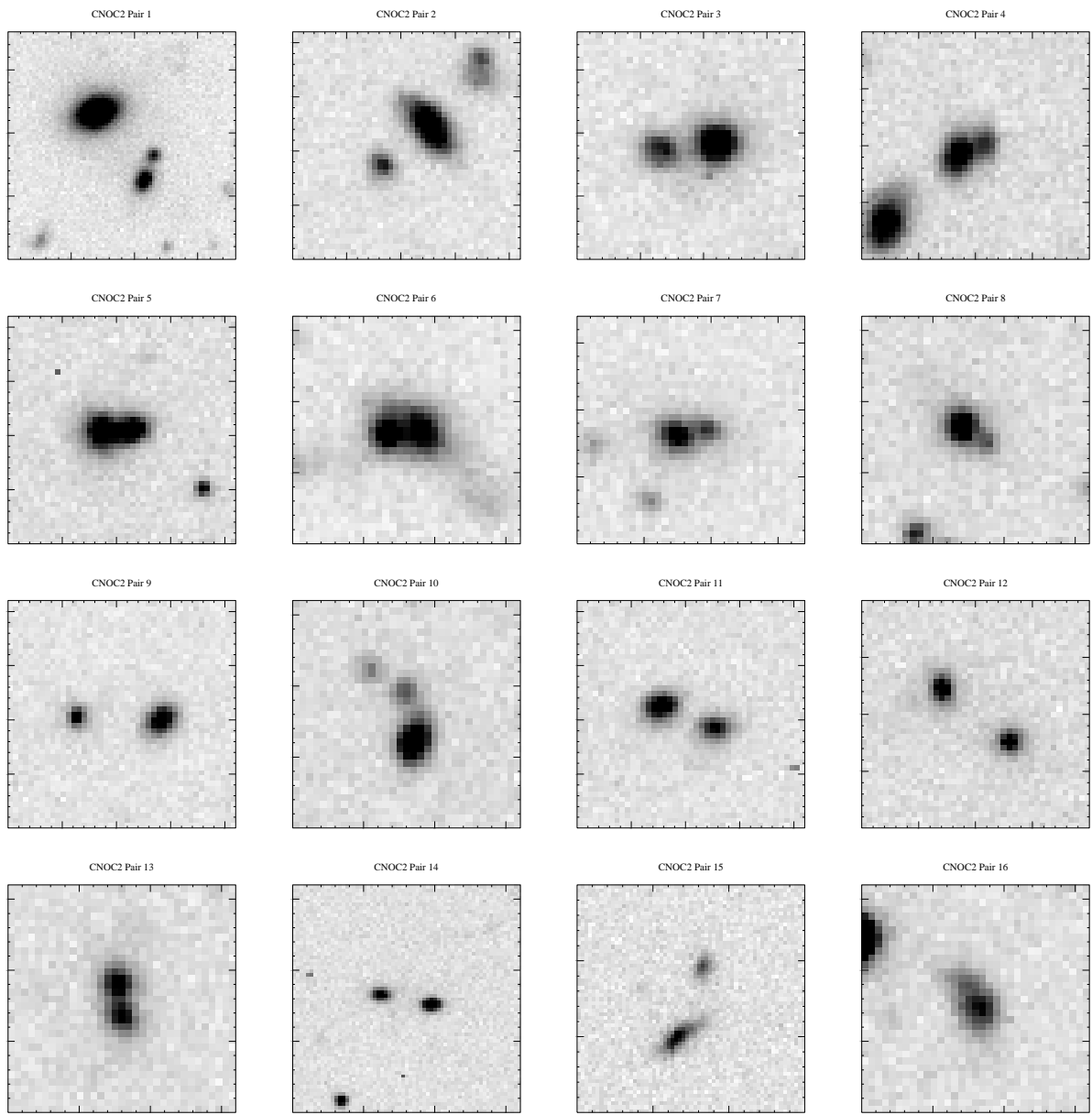}
\caption[Image Mosaic of Close Dynamical Pairs]
{A mosaic of images is given for the 44
close (5 \hkpc~$ < r_p \leq 20$ \hkpc) dynamical 
($\Delta V < 500$ km/s) pairs or triples satisfying the criterion 
used for computing pair statistics.  
These $R_C-$band images were obtained using the CFHT MOS.  
Each image is 50~\hkpc~ on a side, corresponding to 
typical angular sizes of $\sim$ 20\arcsec.  
For images which contain more than two objects, note that 
(a) the image is centered on the pair, (b) the faintest 
galaxies may fall below the flux limit, and (c) other objects 
may be stars or foreground/background galaxies.
\label{cnoc2mr:figim}
}
\end{figure}

\setcounter{figure}{4}
\begin{figure}
\plotone{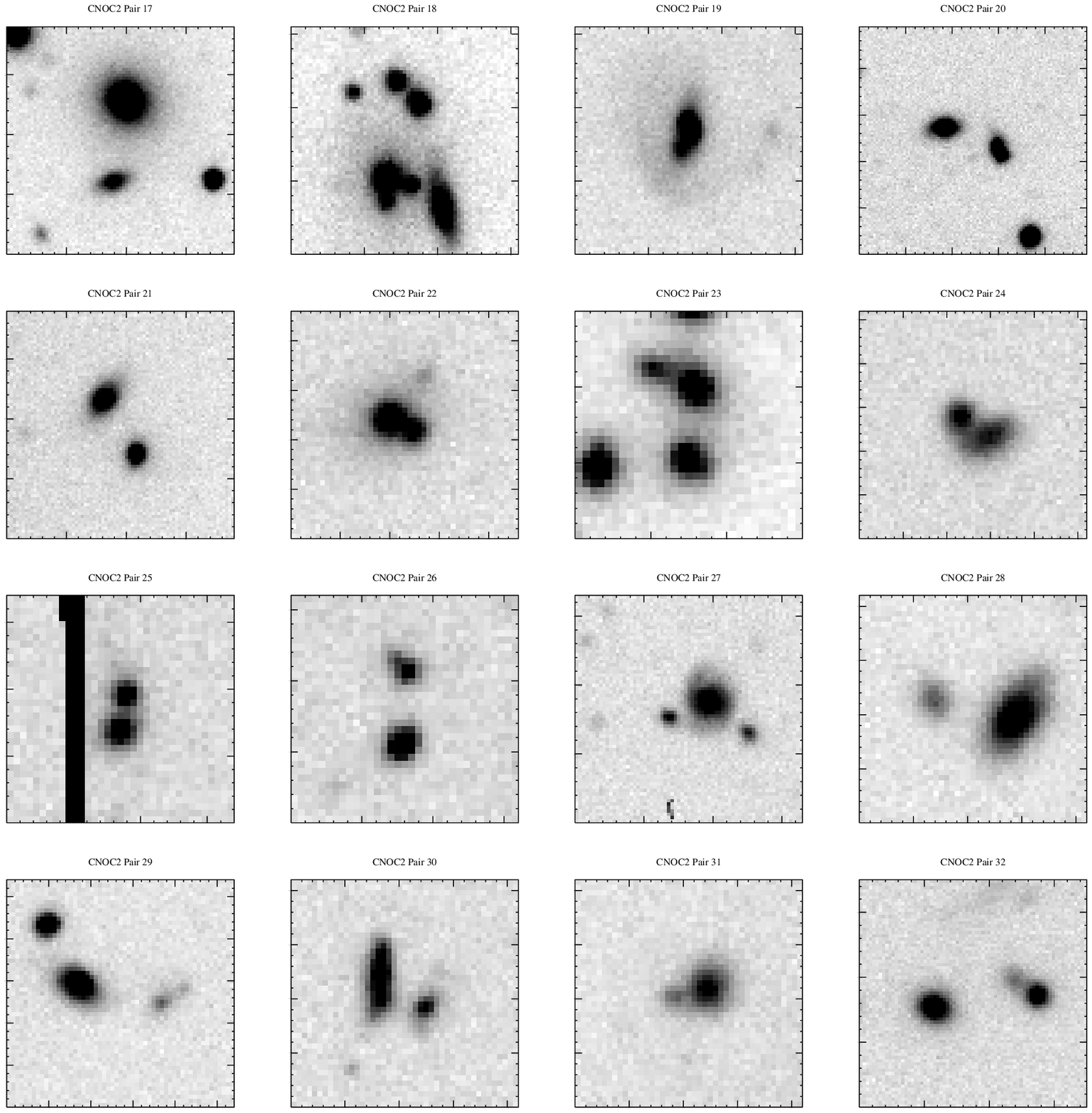}
\caption[]
{
Continued
}
\end{figure}

\setcounter{figure}{4}
\begin{figure}
\plotone{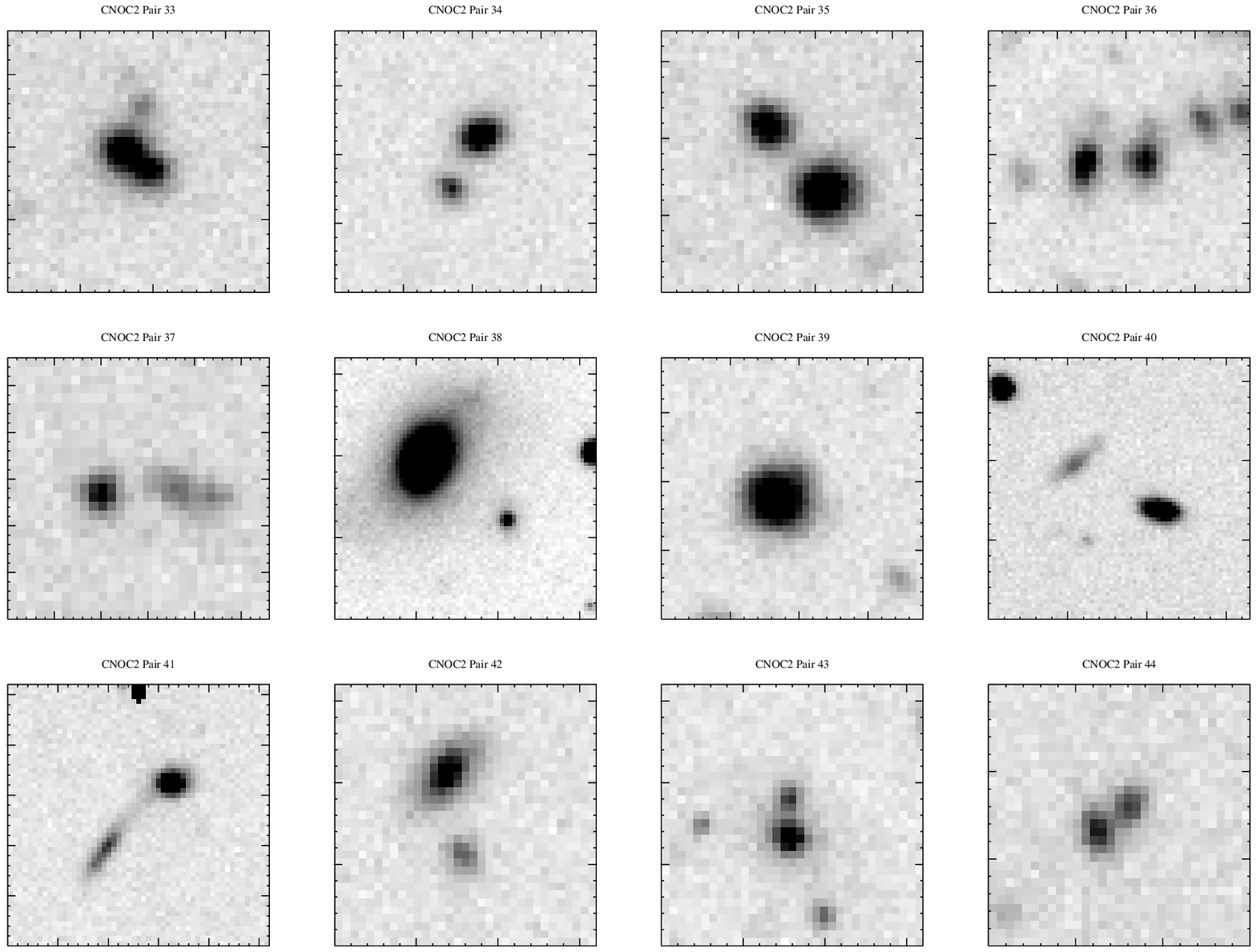}
\caption[]
{
Continued
}
\end{figure}

\begin{figure}
\plotone{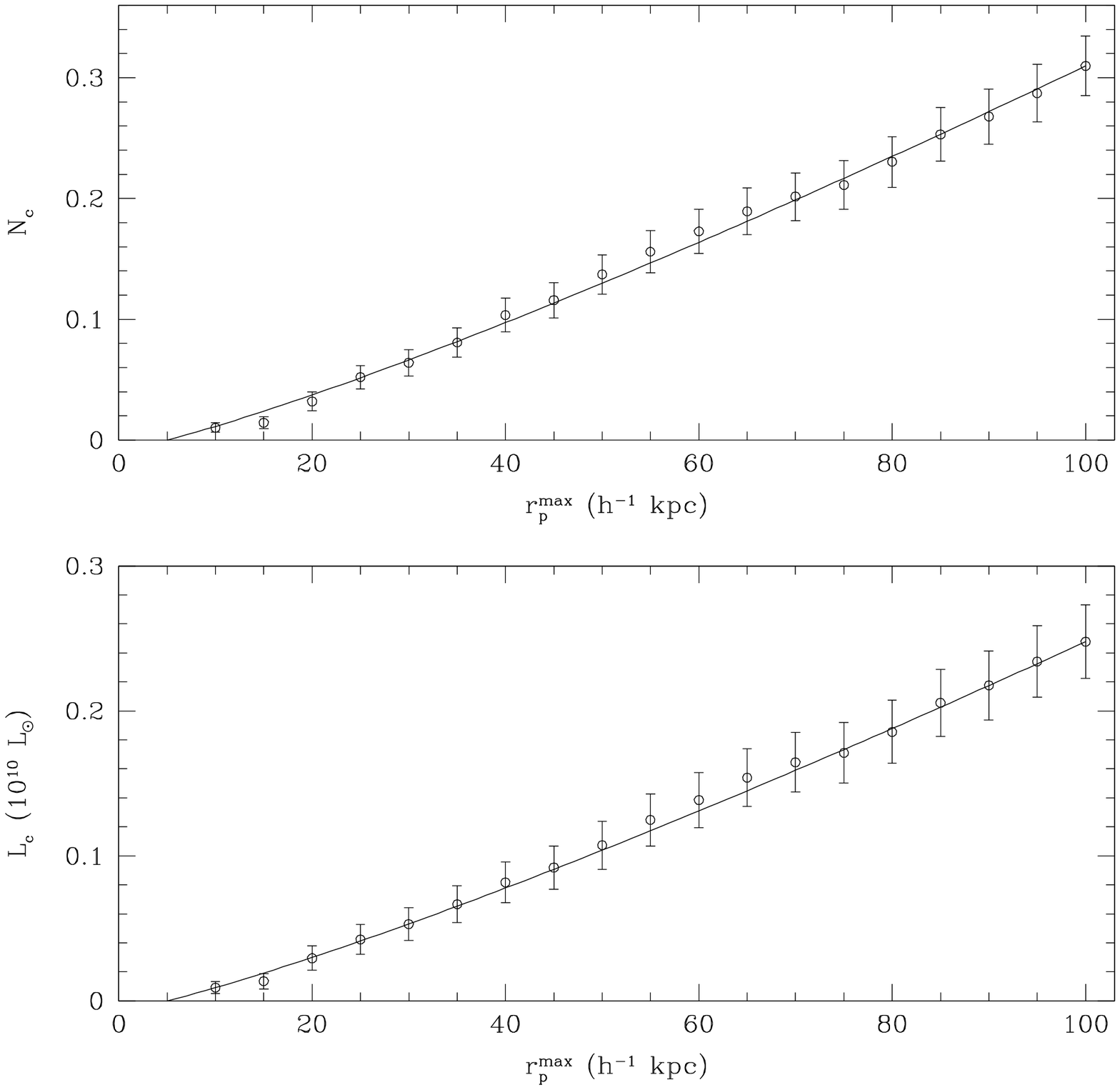}
\caption[Dependence on \rpmax]
{
Pair statistics are computed for $\Delta v \leq 500$ km/s, 
for a range of maximum projected separations ($\rpmax$).  
A minimum projected separation of $r_p=5~$\hkpc~is applied 
in each case.  
Error bars are computed using the Jackknife technique.
Both $N_c$ and $L_c$ are cumulative statistics; hence, 
measurements in successive bins are not independent.  
The line in each panel follows the $r_p^{3-\gamma}$ 
dependence that results from integration over the 
galaxy two-point correlation function.  Taking $\gamma$ = 1.8 
and requiring a match with the data at $r_p = 100~$\hkpc, 
good agreement is found at all separations.    
\label{cnoc2mr:figrp}}
\end{figure}

\begin{figure}
\plotone{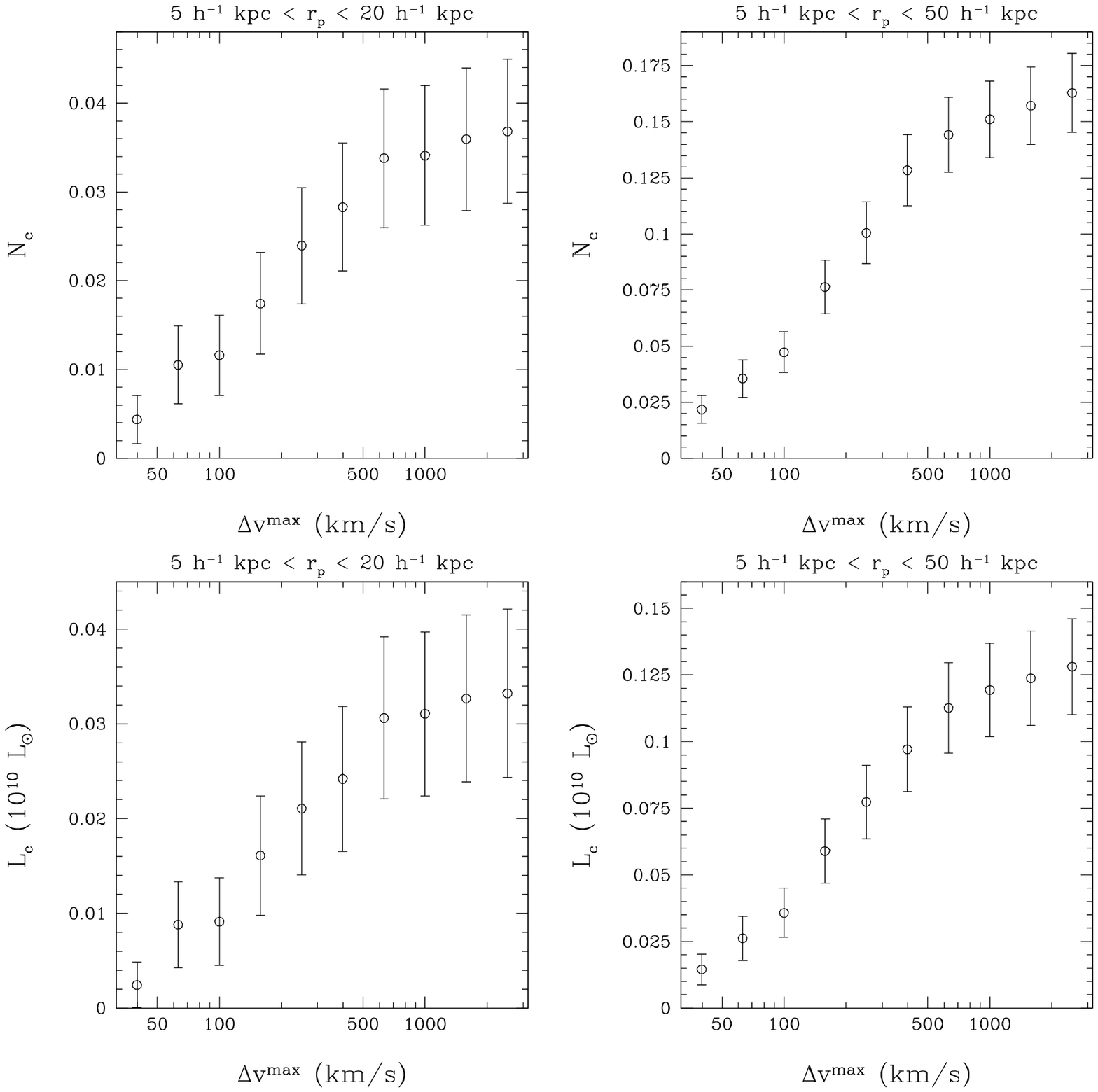}
\caption[Dependence on $\Delta \vmax$]
{
Pair statistics are computed for a range in $\Delta \vmax$,
for both $\rpmax = 20~$\hkpc~(left panels) 
and $\rpmax = 50~$\hkpc~ (right panels).  
Error bars are computed using the Jackknife technique.
Both $N_c$ and $L_c$ are cumulative statistics; hence, 
measurements in successive bins are not independent.  
\label{cnoc2mr:figrl}}
\end{figure}

\begin{figure}
\plotone{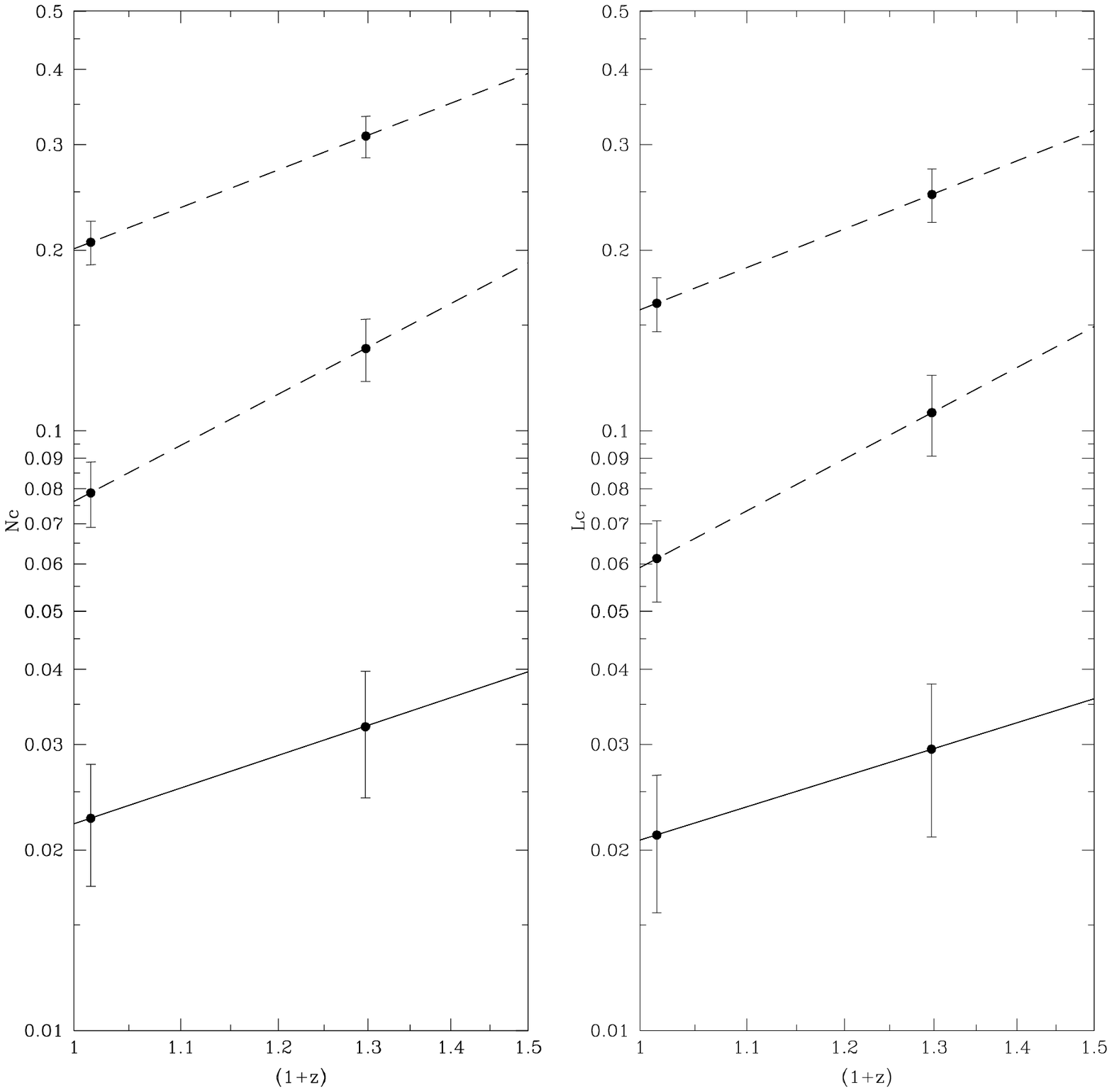}
\caption[Redshift Evolution of CNOC2 Pair Statistics]
{
Pair statistics ($N_c$ and $L_c$) are given for SSRS2 
($<$$z$$>$=0.015) and CNOC2 ($<$$z$$>$=0.30).  
Measurements are given for 
$\rpmax$=20~\hkpc~(bottom), 
$\rpmax$=50~\hkpc~(middle), and 
$\rpmax$=100~\hkpc~(top).
Best fit lines indicate evolution varying as $(1+z)^m$.
For $\rpmax$=(20,50,100)~\hkpc, values of $m$ used are (1.44,2.26,1.66) 
for $N_c$ and (1.34,2.28,1.70) for $L_c$.
Increases in both $N_c$ and $L_c$ with redshift are seen for all three 
choices of $\rpmax$.
\label{cnoc2mr:figmrate1}}
\end{figure}

%% If you are not including electonic art with your submission, you may
%% mark up your captions using the \figcaption command. See the 
%% User Guide for details.
%%
%% No more than seven \figcaption commands are allowed per page, 
%% so if you have more than seven captions, insert a \clearpage 
%% after every seventh one. 

%% Tables should be submitted one per page, so put a \clearpage before
%% each one.

%% Two options are available to the author for producing tables:  the
%% deluxetable environment provided by the AASTeX package or the LaTeX
%% table environment.  Use of deluxetable is preferred.
%%

%% Three table samples follow, two marked up in the deluxetable environment,
%% one marked up as a LaTeX table.

%% In this first example, note that the \tabletypesize{}
%% command has been used to reduce the font size of the table.
%% Note also that the \label command needs to be placed 
%% inside the \tablecaption.

\clearpage
\begin{table}[h]
\caption{CNOC2 0223+00 Close Pairs \label{cnoc2mr:tabcp02}}
\begin{center}
\begin{tabular}[t]{|c|c|r|r|c|c|c|}
\hline
Pair&CNOC2 ID's&$r_p$~~~~~&$\Delta v$~~~&RA&DEC&$\bar{z}$\\
ID&(Galaxy A,B)&($h^{-1}$kpc)&(km/s)&[2000.0]&[2000.0]&\\
\hline
% Following line inserts 0223cp.tex here
%\input{0223cp}
1&100810,100778&18.3&103&02:26:53.2&+00:02:44&0.13339\\
2&140091,140075&13.2&232&02:25:23.0&+00:07:11&0.26693\\
3&130708,130707&12.7&133&02:25:09.2&+00:18:05&0.35090\\
4&030307,030309&6.0&162&02:25:49.2&+00:30:30&0.29829\\
5&191878,191879&6.0&318&02:23:28.3&-00:03:52&0.27009\\
6&131533,131534&6.7&273&02:25:26.4&+00:21:44&0.40391\\
7&121099,121101&6.6&63&02:26:17.1&+00:12:47&0.38407\\
8&041013,041009&5.6&154&02:26:00.5&+00:41:12&0.41931\\
9&050675,050679&19.1&189&02:25:55.5&+00:47:14&0.27099\\
10&031099,031118&11.5&264&02:25:49.0&+00:33:57&0.40769\\
11&140119,140109&12.8&76&02:25:35.4&+00:07:19&0.26747\\
12&030571,030555&19.4&189&02:25:58.9&+00:31:37&0.30031\\
13&040997,040989&7.4&62&02:26:17.4&+00:41:12&0.39882\\
14&160751,160754&11.1&76&02:25:02.5&+00:01:15&0.14997\\
15&010594,010632&16.6&9&02:26:03.6&+00:17:13&0.26799\\
16&040345,040350&5.5&101&02:26:13.4&+00:38:22&0.39666\\
\hline
\end{tabular}
\end{center}
\end{table}

\begin{table}[h]
\caption{CNOC2 0920+37 Close Pairs \label{cnoc2mr:tabcp09}}
\begin{center}
\begin{tabular}[t]{|c|c|r|r|c|c|c|}
\hline
Pair&CNOC2 ID's&$r_p$~~~~~&$\Delta v$~~~&RA&DEC&$\bar{z}$\\
ID&(Galaxy A,B)&($h^{-1}$kpc)&(km/s)&[2000.0]&[2000.0]&\\
\hline
% Following line inserts 0920cp.tex here
%\input{0920cp}
17&010722,010688&18.2&423&09:24:03.2&+37:04:47&0.19171\\
18&061308,061331&18.1&2&09:24:01.2&+37:44:18&0.24652\\
19&081315,081301&5.4&284&09:24:39.6&+37:08:16&0.24645\\
20&050264,050258&13.1&193&09:23:42.8&+37:31:42&0.13640\\
21&010879,010860&13.8&360&09:24:08.0&+37:05:39&0.19029\\
22&131238,131230&6.0&50&09:23:12.6&+37:07:05&0.39017\\
23&172734,172762&16.6&437&09:22:26.9&+36:42:45&0.47497\\
24&020557,020566&7.4&29&09:23:30.7&+37:11:19&0.32471\\
25&160826,160832&8.7&198&09:22:27.9&+36:46:52&0.39208\\
26&191850,191890&17.3&10&09:21:06.9&+36:41:14&0.44085\\
\hline
\end{tabular}
\end{center}
\end{table}

\begin{table}[h]
\caption{CNOC2 1447+09 Close Pairs \label{cnoc2mr:tabcp14}}
\begin{center}
\begin{tabular}[t]{|c|c|r|r|c|c|c|}
\hline
Pair&CNOC2 ID's&$r_p$~~~~~&$\Delta v$~~~&RA&DEC&$\bar{z}$\\
ID&(Galaxy A,B)&($h^{-1}$kpc)&(km/s)&[2000.0]&[2000.0]&\\
\hline
% Following line inserts 1447cp.tex here
%\input{1447cp}
27&101867,101865&9.6&5&14:50:25.4&+08:56:25&0.14608\\
28&140046,140054&18.1&404&14:48:56.1&+08:56:44&0.26193\\
29&162151,162141&18.4&43&14:48:22.0&+08:56:32&0.19365\\
30&120368,120356&12.1&212&14:49:45.8&+08:57:30&0.27244\\
31&110843,110838&6.5&172&14:49:27.0&+08:52:10&0.26965\\
32&011270,011283&19.5&94&14:49:33.1&+09:10:53&0.21406\\
33&051065,051056&5.9&138&14:49:55.1&+09:38:43&0.34926\\
34&161997,161985&12.0&111&14:48:12.9&+08:55:53&0.32429\\
35&082155,082172&16.2&171&14:50:05.9&+09:11:54&0.36420\\
36&091998,091999&11.7&41&14:50:09.9&+09:04:35&0.32469\\
37&040704,040709&15.0&44&14:49:38.9&+09:29:30&0.51024\\
\hline
\end{tabular}
\end{center}
\end{table}

\begin{table}[h]
\caption{CNOC2 2148-05 Close Pairs \label{cnoc2mr:tabcp21}}
\begin{center}
\begin{tabular}[t]{|c|c|r|r|c|c|c|}
\hline
Pair&CNOC2 ID's&$r_p$~~~~~&$\Delta v$~~~&RA&DEC&$\bar{z}$\\
ID&(Galaxy A,B)&($h^{-1}$kpc)&(km/s)&[2000.0]&[2000.0]&\\
\hline
% Following line inserts 2148cp.tex here
%\input{2148cp}
38&070479,070450&19.8&249&21:51:11.5&-04:47:59&0.15447\\
39&180896,180907&6.6&107&21:49:18.5&-05:58:46&0.31239\\
40&141675,141692&18.4&115&21:50:56.9&-05:38:54&0.14462\\
41&031553,031529&16.9&48&21:51:12.7&-05:17:46&0.19767\\
42&071349,071332&16.8&143&21:51:22.0&-04:45:16&0.40839\\
43&131817,131825&7.9&6&21:50:36.8&-05:29:30&0.42589\\
44&162261,162264&8.1&125&21:50:16.5&-05:45:46&0.50631\\
\hline
\end{tabular}
\end{center}
\end{table}

\begin{table}[h]
\caption{CNOC2 Pair Statistics \label{cnoc2mr:tabstats}}
\begin{center}
\begin{tabular}[t]{|c|c|c|c|c|c|}
\hline
Sample&N&N$_{{\rm comp}}$&$\bar{z}$&$N_c$&$L_c (10^{10} h^2 \lsun)$\\
\hline
% Following line inserts cnoc2table.tex here
%\input{cnoc2table}
0223&  1150&  32& 0.309& 0.0538$\pm$ 0.0205& 0.0451$\pm$ 0.0215\\
0920&  1094&  20& 0.301& 0.0252$\pm$ 0.0113& 0.0268$\pm$ 0.0134\\
1447&   961&  22& 0.301& 0.0363$\pm$ 0.0166& 0.0311$\pm$ 0.0173\\
2148&   979&  14& 0.277& 0.0280$\pm$ 0.0176& 0.0213$\pm$ 0.0180\\
CNOC2&  4184&  88& 0.297& 0.0321$\pm$ 0.0077& 0.0294$\pm$ 0.0084\\
\hline
\end{tabular}
\end{center}
\end{table}

\begin{table}[h]
\caption{Pair Statistics Using $\rpmax$ = 50 \hkpc\label{cnoc2mr:tabstats50}}
\begin{center}
\begin{tabular}[t]{|c|c|c|c|c|c|}
\hline
Sample&N&N$_{{\rm comp}}$&$\bar{z}$&$N_c$&$L_c (10^{10} h^2 \lsun)$\\
\hline
% Following line inserts stats50.tex here
%\input{stats50}
SSRS2&  4769& 276& 0.015& 0.0788$\pm$ 0.0099& 0.0613$\pm$ 0.0095\\
CNOC2&  4184& 389& 0.298& 0.1371$\pm$ 0.0163& 0.1072$\pm$ 0.0166\\
\hline
\end{tabular}
\end{center}
\end{table}

\begin{table}[h]
\caption{CNOC2 Pair Statistics for Various Choices of $M_2(B)$
\label{cnoc2mr:tabm2}}
\begin{center}
\begin{tabular}[t]{|c|c|c|}
\hline
$M_2$&$N_c$&$L_c (10^{10} h^2 \lsun)$\\
\hline
% Following line inserts cnoc2m2table.tex here
%\input{cnoc2m2table}
 -19.0& 0.0120$\pm$ 0.0029& 0.0184$\pm$ 0.0052\\
 -18.9& 0.0137$\pm$ 0.0033& 0.0197$\pm$ 0.0056\\
 -18.8& 0.0154$\pm$ 0.0037& 0.0210$\pm$ 0.0060\\
 -18.7& 0.0172$\pm$ 0.0041& 0.0223$\pm$ 0.0064\\
 -18.6& 0.0192$\pm$ 0.0046& 0.0235$\pm$ 0.0067\\
 -18.5& 0.0211$\pm$ 0.0050& 0.0246$\pm$ 0.0070\\
 -18.4& 0.0232$\pm$ 0.0055& 0.0257$\pm$ 0.0073\\
 -18.3& 0.0253$\pm$ 0.0060& 0.0267$\pm$ 0.0076\\
 -18.2& 0.0275$\pm$ 0.0066& 0.0277$\pm$ 0.0079\\
 -18.1& 0.0298$\pm$ 0.0071& 0.0286$\pm$ 0.0082\\
 -18.0& 0.0321$\pm$ 0.0077& 0.0294$\pm$ 0.0084\\
 -17.9& 0.0345$\pm$ 0.0082& 0.0302$\pm$ 0.0086\\
 -17.8& 0.0369$\pm$ 0.0088& 0.0310$\pm$ 0.0088\\
 -17.7& 0.0394$\pm$ 0.0094& 0.0317$\pm$ 0.0090\\
 -17.6& 0.0419$\pm$ 0.0100& 0.0323$\pm$ 0.0092\\
 -17.5& 0.0445$\pm$ 0.0106& 0.0329$\pm$ 0.0094\\
 -17.4& 0.0471$\pm$ 0.0112& 0.0334$\pm$ 0.0096\\
 -17.3& 0.0498$\pm$ 0.0119& 0.0340$\pm$ 0.0097\\
 -17.2& 0.0525$\pm$ 0.0125& 0.0344$\pm$ 0.0098\\
 -17.1& 0.0553$\pm$ 0.0132& 0.0349$\pm$ 0.0100\\
 -17.0& 0.0581$\pm$ 0.0139& 0.0353$\pm$ 0.0101\\
\hline
\end{tabular}
\end{center}
\end{table}

\end{document}